# An Improved Sphere-Packing Bound for Finite-Length Codes on Symmetric Memoryless Channels

Gil Wiechman      Igal Sason

October 29, 2018

## Abstract

This paper derives an improved sphere-packing (ISP) bound for finite-length codes whose transmission takes place over symmetric memoryless channels. We first review classical results, i.e., the 1959 sphere-packing (SP59) bound of Shannon for the Gaussian channel, and the 1967 sphere-packing (SP67) bound of Shannon et al. for discrete memoryless channels. A recent improvement on the SP67 bound, as suggested by Valembois and Fossorier, is also discussed. These concepts are used for the derivation of a new lower bound on the decoding error probability (referred to as the ISP bound) which is uniformly tighter than the SP67 bound and its recent improved version. The ISP bound is applicable to symmetric memoryless channels, and some of its applications are exemplified. Its tightness is studied by comparing it with bounds on the ML decoding error probability, and computer simulations of iteratively decoded turbo-like codes. The paper also presents a technique which performs the entire calculation of the SP59 bound in the logarithmic domain, thus facilitating the exact calculation of this bound for moderate to large block lengths without the need for the asymptotic approximations provided by Shannon.

## I. Introduction

One of Shannon's favorite research topics was the theoretical study of the fundamental performance limits of long block codes. During the fifties and sixties, this research work attracted Shannon and his colleagues at MIT and Bell Labs; the contributions which came out of this work were published by Shannon et al. (see, e.g., the collected papers of Shannon in [26] and the book of Gallager [12]). An overview of these results and their impact was addressed by Berlekamp [2].

The introduction of turbo-like codes, which closely approach the Shannon capacity limit with moderate block lengths and feasible decoding complexity, stirred up new interest in studying the limits of code performance as a function of the block length (see, e.g., [9], [14], [15], [17], [23], [29], [35], [37]).

Following this direction of research, this paper is aimed to contribute to the study of the fundamental performance limitations of finite-length codes whose transmission takes place over an arbitrary symmetric memoryless channel, and also to study the fundamental tradeoff between the performance and block length of these codes. This study is facilitated by theoretical bounds, and is also compared with practical results which are obtained by modern coding techniques and sub-optimal decoding algorithms. In this respect, the reader is referred to a recent and comprehensive tutorial paper by Costello and Forney [3] which traces the evolution of channel coding techniques, and also addresses the significant contribution of error-correcting codes in improving the tradeoff between performance, block length (delay) and complexity for practical applications.

The 1959 sphere-packing (SP59) bound of Shannon [24] serves for the evaluation of the performance limits of block codes whose transmission takes place over an AWGN channel. This lower bound on the decoding error probability is expressed in terms of the block length and rate of the code; however, it does not take into account the modulation used, but only assumes that the signals are of equal energy. It is often used as a reference for quantifying the sub-optimality of error-correcting codes associated with their decoding algorithms.

The 1967 sphere-packing (SP67) bound, derived by Shannon, Gallager and Berlekamp [25], provides a lower bound on the decoding error probability of block codes as a function of their block length and code rate, and it applies to arbitrary discrete memoryless channels. Like the random coding bound of Gallager [11], the SP67 bound decays to zero exponentially with the block length for all rates below the channel capacity. Further, the error exponent of the SP67 bound is tight at the portion of the rate region between the critical rate ($R_c$) and the channel capacity; for all rates in this range, the error exponents of the SP67 and the random coding bounds coincide (see [25, Part 1]).

In spite of its exponential behavior, the SP67 bound appears to be loose for codes of small to moderate block lengths. This weakness is due to the original focus in [25] on asymptotic analysis. In their paper [35], Valembois and Fossorier revisited the SP67 bound in order to improve its tightness for finite-length block codes (especially, for codes of short to moderate block lengths), and also extended its validity to memoryless continuous-output channels (e.g., the binary-input AWGN channel). The remarkable improvement of their bound over the classical SP67 bound was exemplified in [35]; moreover, it provides an interesting alternative to the SP59 bound which is particularized for the AWGN channel [24].

In this work, we derive an improved sphere-packing bound (referred to as the *ISP bound*) which further enhances the tightness of the bounding technique in [25], especially for codes of short to moderate block lengths; this new bound is valid for all symmetric memoryless channels.

The paper is structured as follows: Section II reviews the concepts used in the derivation of the SP67 bound [25, Part 1], and its recent improvements in [35] which are especially effective for codes of short to moderate block lengths. In Section III, we derive the ISP bound which further enhances the tightness of the bound in [35] for symmetric memoryless channels; the derivation of this bound relies on concepts and notation presented in Section II. Section IV starts by reviewing the SP59 bound of Shannon [24], and presenting the numerical algorithm used in [35] for calculating this bound. The numerical instability of this algorithm for codes of moderate to large block lengths motivates the derivation of an alternative algorithm in Section IV which facilitates the exact calculation of the SP59 bound, irrespectively of the block length. Section V provides numerical results which serve to compare the tightness of the ISP bound, derived in Section III, with the SP59 bound of Shannon [24] and the recent sphere-packing bound in [35]. The tightness of the ISP bound is exemplified in Section V for M-ary phase-shift-keying (PSK) block coded modulation schemes whose transmission takes place over the AWGN channel, and also for the binary erasure channel (BEC). Additionally, Section V applies the sphere-packing bounds to give lower bounds on the block length required to achieve a required performance on a given channel. These lower bounds are compared with the performance of some practically decodable codes which are presented in recent works. We conclude our discussion in Section VI. Technical calculations are relegated to the appendices.

## II. The 1967 Sphere-Packing Bound and Improvements

In this section, we outline the derivation of the SP67 bound. We then survey the improvements to this bound, as suggested in [35], which also extend the validity of the improved bound to memoryless discrete-input continuous-output channels. This review serves as a preparatory stage for presenting an improved sphere-packing bound in the next section; the new bound further enhances the tightness of the sphere-packing bounding technique for finite-length codes whose transmission takes place over symmetric memoryless channels. For a comprehensive tutorial review of sphere-packing bounds, the reader is referred to [23, Chapter 5]. Due to the strong relevance of the sphere-packing bounds to the analysis in this paper, we note that the two "Information and Control" papers related to the SP67 bound [25] and the paper related to the SP59 bound [24] are also published in the book which consists of all the papers of Shannon [26].

### A. The 1967 Sphere-Packing Bound

Let us consider a block code $\mathcal{C}$ which consists of $M$ codewords each of length $N$, and denote its codewords by $\mathbf{x}_1, \ldots, \mathbf{x}_M$. Assume that $\mathcal{C}$ is transmitted over a discrete memoryless channel (DMC) and decoded by a *list decoder*; for each received sequence $\mathbf{y}$, the decoder outputs a list of at most $L$ integers belonging to the set $\{1, 2, \ldots, M\}$ which correspond to the indices of the codewords. A list decoding error is declared if the index of the transmitted codeword does not appear in the list. List decoding, originally introduced by Elias [10] and Wozencraft [38], signifies an important class of decoding algorithms. In [25], the authors derive a lower bound on the decoding error probability of an arbitrary block code with $M$ codewords of length $N$; the bound applies to an arbitrary list decoder where the size of the list is limited to $L$. The particular case where $L = 1$ clearly provides a lower bound on the decoding error probability under maximum-likelihood (ML) decoding.

Let $\mathcal{Y}_m$ denote the set of output sequences $\mathbf{y}$ for which message $m$ is on the decoding list, and define $P_m(\mathbf{y}) \triangleq \Pr(\mathbf{y}|\mathbf{x}_m)$. The probability of list decoding error when message $m$ is sent over the channel is given by

$$P_{\mathrm{e},m} = \sum_{y \in \mathcal{Y}_m^{\mathrm{c}}} P_m(\mathbf{y}) \tag{1}$$

where the superscript 'c' stands for the complementary set. For the block code and list decoder under consideration, let $P_{\mathrm{e,max}}$ designate the maximal value of $P_{\mathrm{e},m}$ where $m \in \{1, 2, \ldots, M\}$. Assuming that all the codewords are equally likely to be transmitted, the average decoding error probability is given by

$$P_{\mathrm{e}} = \frac{1}{M} \sum_{m=1}^{M} P_{\mathrm{e},m}.$$

Referring to a list decoder of size at most $L$, the code rate (in nats per channel use) is defined as $R \triangleq \frac{\ln\left(\frac{M}{L}\right)}{N}$.

The derivation of the SP67 bound is divided in [25, Part 1] into three main steps. The first step refers to the derivation of upper and lower bounds on the error probability of a code consisting of two codewords only. These bounds are given by the following theorem:

*Theorem 2.1 (Upper and Lower Bounds on the Pairwise Error Probability):* [25, Theorem 5]. Let $P_1$ and $P_2$ be two probability assignments defined over a discrete set of sequences, $\mathcal{Y}_1$ and $\mathcal{Y}_2 = \mathcal{Y}_1^{\mathrm{c}}$ be (disjoint) decision regions for these sequences, $P_{\mathrm{e},1}$ and $P_{\mathrm{e},2}$ be given by (1), and assume that $P_1(\mathbf{y})P_2(\mathbf{y}) \neq 0$ for at least one sequence $\mathbf{y}$. Then, for all $s \in (0, 1)$

$$P_{\mathrm{e},1} > \frac{1}{4} \exp\left( \mu(s) - s\mu'(s) - s\sqrt{2\mu''(s)} \right) \tag{2}$$

or

$$P_{\mathrm{e},2} > \frac{1}{4} \exp\left( \mu(s) + (1-s)\mu'(s) - (1-s)\sqrt{2\mu''(s)} \right) \tag{3}$$

where

$$\mu(s) \triangleq \ln\left( \sum_{\mathbf{y}} P_1(\mathbf{y})^{1-s} P_2(\mathbf{y})^s \right) \qquad 0 < s < 1. \tag{4}$$

Furthermore, for an appropriate choice of the decision regions $\mathcal{Y}_1$ and $\mathcal{Y}_2$, the following upper bounds hold:

$$P_{\mathrm{e},1} \leq \exp\left( \mu(s) - s\mu'(s) \right) \tag{5}$$

and

$$P_{\mathrm{e},2} \leq \exp\left( \mu(s) + (1-s)\mu'(s) \right). \tag{6}$$

The function $\mu$ is non-positive and convex over the interval $(0, 1)$. The convexity of $\mu$ is strict unless $\frac{P_1(\mathbf{y})}{P_2(\mathbf{y})}$ is constant over all the sequences $\mathbf{y}$ for which $P_1(\mathbf{y})P_2(\mathbf{y}) \neq 0$. Moreover, the function $\mu$ is strictly negative over the interval $(0, 1)$ unless $P_1(\mathbf{y}) = P_2(\mathbf{y})$ for all $\mathbf{y}$.

*Proof:* A full proof of Theorem 2.1 is given in [25, Section III]. In the following, we present a brief outline of the proof which serves to emphasize the parallelism between Theorem 2.1 and the first part of the derivation of the ISP bound in Section III. To this end, let us define the log-likelihood ratio (LLR) as

$$D(\mathbf{y}) \triangleq \ln\left( \frac{P_2(\mathbf{y})}{P_1(\mathbf{y})} \right) \tag{7}$$

and the probability distribution

$$Q_s(\mathbf{y}) \triangleq \frac{P_1(\mathbf{y})^{1-s} P_2(\mathbf{y})^s}{\sum_{\mathbf{y}'} P_1(\mathbf{y}')^{1-s} P_2(\mathbf{y}')^s}, \quad 0 < s < 1. \tag{8}$$

It is simple to show (see [25]) that for all $0 < s < 1$, the first and second derivatives of $\mu$ in (4) are equal to the statistical expectation and variance of the LLR, respectively, taken with respect to (w.r.t.) the probability distribution $Q_s$ in (8). This gives the following equalities:

$$\mu'(s) = \mathbb{E}_{Q_s}\big(D(\mathbf{y})\big) \tag{9}$$

$$\mu''(s) = \text{Var}_{Q_s}\big(D(\mathbf{y})\big) \tag{10}$$

$$P_1(\mathbf{y}) = \exp\big(\mu(s) - sD(\mathbf{y})\big)\, Q_s(\mathbf{y}) \tag{11}$$

$$P_2(\mathbf{y}) = \exp\big(\mu(s) + (1-s)D(\mathbf{y})\big)\, Q_s(\mathbf{y})\,. \tag{12}$$

where equalities (11) and (12) follow easily from (4), (7) and (8). For every $0 < s < 1$, we further define the set of sequences

$$\mathcal{Y}_s \triangleq \Big\{ \mathbf{y} \in \mathcal{Y} : |D(\mathbf{y}) - \mu'(s)| \le \sqrt{2\mu''(s)} \,\Big\}. \tag{13}$$

For any choice of a decision region $\mathcal{Y}_1$, the conditional error probability given that the first message was transmitted satisfies

$$
\begin{aligned}
P_{\text{e},1} &= \sum_{\mathbf{y} \in \mathcal{Y}_1^c} P_1(\mathbf{y}) \\
&\ge \sum_{\mathbf{y} \in \mathcal{Y}_1^c \bigcap \mathcal{Y}_s} P_1(\mathbf{y}) \\
&\stackrel{(a)}{=} \sum_{\mathbf{y} \in \mathcal{Y}_1^c \bigcap \mathcal{Y}_s} \exp\Big(\mu(s) - sD(\mathbf{y})\Big)\, Q_s(\mathbf{y}) \\
&\stackrel{(b)}{\ge} \exp\Big(\mu(s) - s\mu'(s) - s\,\sqrt{2\mu''(s)}\Big) \sum_{\mathbf{y} \in \mathcal{Y}_1^c \bigcap \mathcal{Y}_s} Q_s(\mathbf{y})
\end{aligned}
\tag{14}
$$

where $(a)$ follows from (11) and $(b)$ relies on the definition of $\mathcal{Y}_s$ in (13). Using similar arguments and relying on (12), we also get that

$$P_{\text{e},2} \ge \exp\Big(\mu(s) + (1-s)\mu'(s) - (1-s)\,\sqrt{2\mu''(s)}\Big) \sum_{\mathbf{y} \in \mathcal{Y}_2^c \bigcap \mathcal{Y}_s} Q_s(\mathbf{y})\,. \tag{15}$$

Since $\mathcal{Y}_1$ and $\mathcal{Y}_2$ form a partition of the observation space, we have that

$$\sum_{\mathbf{y} \in \mathcal{Y}_1 \bigcap \mathcal{Y}_s} Q_s(\mathbf{y}) + \sum_{\mathbf{y} \in \mathcal{Y}_2 \bigcap \mathcal{Y}_s} Q_s(\mathbf{y}) = \sum_{\mathbf{y} \in \mathcal{Y}_s} Q_s(\mathbf{y}) > \frac{1}{2}$$

where the last transition relies on (9) and (10) and follows from Chebychev's inequality. Therefore, at least one of the two sums on the LHS of the expression above must be greater than $\frac{1}{4}$. Substituting this in (14) and (15) completes the proof of the lower bound on the error probability in (2) and (3). The upper bound on the error probability in (5) and (6) is attained by selecting the decision region for the first codeword to be

$$\mathcal{Y}_1 \triangleq \big\{ \mathbf{y} \in \mathcal{Y} : D(\mathbf{y}) < \mu'(s) \big\}$$

and the decision region for the second code as $\mathcal{Y}_2 \triangleq \mathcal{Y}_1^c$. The proof for the upper bounds in in (5) and (6) follows directly from (11), (12) and the particular choice of $\mathcal{Y}_1$ and $\mathcal{Y}_2$ above. $\blacksquare$

The initial motivation given for Theorem 2.1 is the calculation of lower bounds on the error probability of a two-word code. However, it is valid for any pair of probability assignments $P_1$ and $P_2$ and decision regions $\mathcal{Y}_1$ and $\mathcal{Y}_2$ which form a partition of the observation space.

In the continuation of the derivation of the SP67 bound in [25], this theorem is used in order to control the size of a decision region of a particular codeword without directly referring to the other codewords. To this end, an arbitrary probability tilting measure $f_N$ is introduced in [25] over all $N$-length sequences of channel outputs, requiring that it is factorized in the form

$$f_N(\mathbf{y}) = \prod_{n=1}^{N} f(y_n) \tag{16}$$

for an arbitrary output sequence $\mathbf{y} = (y_1, \ldots, y_N)$. The size of the set $\mathcal{Y}_m$ is defined as

$$F(\mathcal{Y}_m) \triangleq \sum_{\mathbf{y} \in \mathcal{Y}_m} f_N(\mathbf{y}). \tag{17}$$

Next, [25] relies on Theorem 2.1 in order to relate the conditional error probability $P_{\mathrm{e},m}$ and $F(\mathcal{Y}_m)$ for fixed composition codes; this is done by associating $\Pr(\cdot|\mathbf{x}_m)$ and $f_N$ with $P_1$ and $P_2$, respectively. Theorem 2.1 is applied to derive a parametric lower bound on the size of the decision region $\mathcal{Y}_m$ or on the conditional error probability $P_{\mathrm{e},m}$. Due to the fact that the list size is limited to $L$, then $\mathbf{y} \in \mathcal{Y}_m$ for at most $L$ indices $m \in \{1, \ldots, M\}$ and hence $\sum_{m=1}^{M} F(\mathcal{Y}_m) \leq L$. Therefore, there exists an index $m$ so that $F(\mathcal{Y}_m) \leq \frac{L}{M}$ and for this unknown value of $m$, one can upper bound the conditional error probability $P_{\mathrm{e},m}$ by

$$P_{\mathrm{e},\max} \triangleq \max_{m \in \{1, \ldots, M\}} P_{\mathrm{e},m}.$$

Using Theorem 2.1, this provides a lower bound on $P_{\mathrm{e},\max}$. Next, the probability assignment $f \triangleq f_s$ is optimized in [25], so as to get the tightest (i.e., maximal) lower bound within this form while considering a code whose composition minimizes the bound (so that the bound holds for all fixed composition codes). A solution for this min-max problem, as provided in [25, Eq. 4.18–4.20], leads to the following theorem which gives a lower bound on the maximal block error probability of an arbitrary fixed composition block code (for a more detailed review of these concepts, see [23, Section 5.3]).

*Theorem 2.2 (Sphere-Packing Bound on the Maximal Decoding Error Probability for Fixed Composition Codes):* [25, Theorem 6]. Let $\mathcal{C}$ be a *fixed composition code* of $M$ codewords and block length $N$. Assume that the transmission of $\mathcal{C}$ takes place over a DMC, and let $P(j|k)$ be the set of transition probabilities characterizing this channel (where $j \in \{0, \ldots, J-1\}$ and $k \in \{0, \ldots, K-1\}$ designate the channel output and input, respectively). For an arbitrary list decoder whose list size is limited to $L$, the *maximal error probability* ($P_{\mathrm{e},\max}$) satisfies

$$P_{\mathrm{e},\max} \geq \exp\left[ -N\left( E_{\mathrm{sp}}\left( R - \frac{\ln 4}{N} - \varepsilon \right) + \sqrt{\frac{8}{N}} \ln\left( \frac{e}{\sqrt{P_{\min}}} \right) + \frac{\ln 4}{N} \right) \right]$$

where $R \triangleq \frac{\ln\left(\frac{M}{L}\right)}{N}$ is the rate of the code, $P_{\min}$ designates the smallest non-zero transition probability of the DMC, the parameter $\varepsilon$ is an arbitrarily small positive number, and the function $E_{\mathrm{sp}}$ is given by

$$E_{\mathrm{sp}}(R) \triangleq \sup_{\rho \geq 0} \left( E_0(\rho) - \rho R \right) \tag{18}$$

$$E_0(\rho) \triangleq \max_{\mathbf{q}} E_0(\rho, \mathbf{q}) \tag{19}$$

$$E_0(\rho, \mathbf{q}) \triangleq -\ln\left( \sum_{j=0}^{J-1} \left[ \sum_{k=0}^{K-1} q_k P(j|k)^{\frac{1}{1+\rho}} \right]^{1+\rho} \right). \tag{20}$$

The maximum in the RHS of (19) is taken over all probability vectors $\mathbf{q} = (q_0, \ldots, q_{K-1})$, i.e., over all $\mathbf{q}$ with $K$ non-negative components summing to 1.

The reason for considering fixed composition codes in [25] is that, in general, the optimal probability distribution $f_s$ may depend on the composition of the codewords through the choice of the parameter $s$ in $(0, 1)$ (see [25, p. 96]).

The next step in the derivation of the SP67 bound is the application of Theorem 2.2 towards the derivation of a lower bound on the maximal block error probability of an arbitrary block code. This is performed by lower bounding the maximal block error probability of the code by the maximal block error probability of its largest fixed composition subcode. Since the number of possible compositions is polynomial in the block length, one can lower bound the rate of the largest fixed composition subcode by $R - O\left(\frac{\ln N}{N}\right)$ where $R$ is the rate of the original code. Clearly, the rate loss caused by considering this subcode vanishes when the block length tends to infinity; however, it loosens the bound for codes of short to moderate block lengths. Finally, the bound on the maximal block error probability is transformed into a bound on the average block error probability by considering an expurgated code which contains half of the codewords of the original code with the lowest decoding error probability. This finally leads to the SP67 bound in [25, Part 1].

*Theorem 2.3 (The 1967 Sphere-Packing Bound for Discrete Memoryless Channels):* [25, Theorem 2]. Let $\mathcal{C}$ be an arbitrary block code whose transmission takes place over a DMC. Assume that the DMC is specified by the set of transition probabilities $P(j|k)$ where $k \in \{0, \ldots, K-1\}$ and $j \in \{0, \ldots, J-1\}$ designate the channel input and output alphabets, respectively. Assume that the code $\mathcal{C}$ forms a set of $M$ codewords of length $N$ (i.e., each codeword is a sequence of $N$ letters from the input alphabet), and consider an arbitrary list decoder where the size of the list is limited to $L$. Then, the *average decoding error probability* of the code $\mathcal{C}$ satisfies

$$P_{\mathrm{e}}(N, M, L) \geq \exp\left\{-N\left[E_{\mathrm{sp}}\left(R - O_1\left(\frac{\ln N}{N}\right)\right) + O_2\left(\frac{1}{\sqrt{N}}\right)\right]\right\}$$

where $R \triangleq \frac{\ln\left(\frac{M}{L}\right)}{N}$ and the error exponent $E_{\mathrm{sp}}(R)$ is introduced in (18). The terms

$$\begin{aligned} O_1\left(\frac{\ln N}{N}\right) &= \frac{\ln 8}{N} + \frac{K \ln N}{N} \\ O_2\left(\frac{1}{\sqrt{N}}\right) &= \sqrt{\frac{8}{N}}\,\ln\left(\frac{e}{\sqrt{P_{\min}}}\right) + \frac{\ln 8}{N} \end{aligned} \tag{21}$$

scale like $\frac{\ln N}{N}$ and the inverse of the square root of $N$, respectively (hence, they both vanish as we let $N$ tend to infinity), and $P_{\min}$ denotes the smallest non-zero transition probability of the DMC.

### B. Recent Improvements on the 1967 Sphere-Packing Bound

In [35], Valembois and Fossorier revisit the derivation of the SP67 bound, focusing this time on finite-length block codes. They present four modifications to the classical derivation in [25] which improve the pre-exponent of the SP67 bound. The new bound derived in [35] is also valid for memoryless channels with discrete input and continuous output (as opposed to the SP67 bound which is only valid for DMCs). It is applied to the binary-input AWGN channel, and is also compared with the SP59 bound which holds for any set of equal energy signals transmitted over the AWGN channel; this comparison shows that the recent bound in [35] provides an interesting alternative to the SP59 bound, especially for high code rates. In this section, we outline the improvements suggested in [35] and present the resulting bound.

The first modification suggested in [35] is the addition of a free parameter in the derivation of the lower bound on the decoding error probability of two-word codes; this free parameter is used in conjunction with Chebychev's inequality, and it is optimized in order to get the tightest bound within this form.

A second improvement presented in [35] is related to a simplification in [25, Part 1] where the inequality $s\sqrt{\mu''(s)} \leq \ln\left(\frac{e}{\sqrt{P_{\min}}}\right)$ is applied. This bound on the second derivative of $\mu$ results in no asymptotic loss, but loosens the bound on the decoding error probability for short to moderate block lengths. By using the exact value of $\mu''$ instead, the tightness of the resulting bound is further improved in [35]. This modification also makes the bound suitable to memoryless channels with continuous output, as it is no longer required that $P_{\min}$ is positive. It should be noted that this causes a small discrepancy in the derivation of the bound; the derivation of a lower bound on the block error probability which is *uniform* over all fixed composition codes relies on finding the composition which minimizes the lower bound. The optimal composition is given in [25, Eq. 4.18, 4.19] for the case where the upper bound on $\mu''$ is applied. In [35], the same composition is used without checking whether it is still the composition which minimizes the lower bound. However, as we see in the next section, for the class of symmetric memoryless channels the value of the bound is independent of the code composition; therefore, the bound of Valembois and Fossorier [35, Theorem 7] (referred to as the VF bound) stays valid. This class of channels includes all memoryless binary-input output-symmetric (MBIOS) channels.

A third improvement in [35] concerns the particular selection of the value of $\rho \geq 0$ which leads to the derivation of Theorem 2.3. In [25], $\rho$ is set to be the value $\tilde{\rho}$ which maximizes the error exponent of the SP67 bound (i.e., the upper bound on the error exponent). This choice emphasizes the similarity between the error exponents of the SP67 bound and the random coding bound, hence proving that the error exponent of the SP67 bound is tight for all rates above the critical rate of the channel. In order to tighten the bound for the finite-length case, [35] chooses the value of $\rho$ to be $\rho^*$ which provides the tightest possible lower bound on the decoding error probability. For rates above the critical rate of the channel, the asymptotic accuracy of the original SP67 bound implies that as the block length tends to infinity, $\tilde{\rho}$ tends to $\rho^*$. However, for codes of finite block length, this simple observation tightens the bound with almost no penalty in the computational complexity of the resulting bound.

The fourth observation made in [35] concerns the final stage in the derivation of the SP67 bound. In order to get a lower bound on the maximal block error probability of an arbitrary block code, the derivation in [25] considers the maximal block error probability of a fixed composition subcode of the original code. In [25], a simple lower bound on the size of the largest fixed composition subcode is given; namely, the size of the largest fixed composition subcode is not less than the size of the entire code divided by the number of possible compositions. Since the number of possible compositions is equal to the number of possible ways to divide $N$ symbols into $K$ types, this value is given by $\binom{N+K-1}{K-1}$. To simplify the final expression of the SP67 bound, [25] applies the upper bound $\binom{N+K-1}{K-1} \leq N^K$. Since this expression is polynomial in the block length $N$, there is no asymptotic loss to the error exponent. However, by using the exact expression for the number of possible compositions, the bound in [35] is tightened for codes of short to moderate block lengths. Applying these four modifications in [35] yields an improved lower bound on the decoding error probability of block codes transmitted over memoryless channels with finite input alphabets. As mentioned above, these modifications also extend the validity of the new bound to memoryless channels with discrete input and continuous output. However, the requirement of a finite input alphabet still remains, as it is required in order to apply the bound to arbitrary block codes, and not only to fixed composition codes. The VF bound [35] is given in the following theorem:

*Theorem 2.4 (Improvement on the 1967 Sphere-Packing Bound for Discrete Memoryless Channels):* [35, Theorem 7]. Under the assumptions and notation used in Theorem 2.3, the *average decoding error probability* satisfies

$$P_{\mathrm{e}}(N, M, L) \geq \exp\left\{-N \widetilde{E}_{\mathrm{sp}}(R, N)\right\}$$

where

$$\widetilde{E}_{\mathrm{sp}}(R, N) \triangleq \sup_{x > \frac{\sqrt{2}}{2}} \left\{ E_0(\rho_x) - \rho_x \left( R - O_1\left(\frac{\ln N}{N}, x\right) \right) + O_2\left(\frac{1}{\sqrt{N}}, x, \rho_x\right) \right\}$$

and

$$O_1\left(\frac{\ln N}{N}, x\right) \triangleq \frac{\ln 8}{N} + \frac{\ln \binom{N+K-1}{K-1}}{N} - \frac{\ln\left(2 - \frac{1}{x^2}\right)}{N} \tag{22}$$

$$O_2\left(\frac{1}{\sqrt{N}}, x, \rho\right) \triangleq x \sqrt{\frac{8}{N} \sum_{k=0}^{K-1} q_{k,\rho} \nu_k^{(2)}(\rho)} + \frac{\ln 8}{N} - \frac{\ln\left(2 - \frac{1}{x^2}\right)}{N}$$

$$\nu_k^{(1)}(\rho) \triangleq \frac{\displaystyle\sum_{j=0}^{J-1} \beta_{j,k,\rho} \ln \frac{\beta_{j,k,\rho}}{P(j|k)}}{\displaystyle\sum_{j=0}^{J-1} \beta_{j,k,\rho}}$$

$$\nu_k^{(2)}(\rho) \triangleq \frac{\displaystyle\sum_{j=0}^{J-1} \beta_{j,k,\rho} \ln^2 \frac{\beta_{j,k,\rho}}{P(j|k)}}{\displaystyle\sum_{j=0}^{J-1} \beta_{j,k,\rho}} - \left[\nu_k^{(1)}(\rho)\right]^2$$

$$\beta_{j,k,\rho} \triangleq P(j|k)^{\frac{1}{1+\rho}} \cdot \left( \sum_{k'=0}^{K-1} q_{k',\rho} P(j|k')^{\frac{1}{1+\rho}} \right)^{\rho}$$

where $\mathbf{q}_\rho \triangleq (q_{1,\rho}, \ldots, q_{K,\rho})$ designates the input distribution which maximizes $E_0(\rho, \mathbf{q})$ in (19), and the parameter $\rho = \rho_x$ is determined by solving the equation

$$R - O_1\left(\frac{\ln N}{N}, x\right) = -\frac{1}{\rho} \sum_{k=0}^{K-1} q_{k,\rho} \nu_k^{(1)}(\rho) + \frac{x}{\rho} \sqrt{\frac{2}{N} \sum_{k=0}^{K-1} q_{k,\rho} \nu_k^{(2)}(\rho)}.$$

For a more detailed review of the improvements suggested in [35], the reader is referred to [23, Section 5.4].

### III. An Improved Sphere-Packing Bound

In this section, we derive an improved lower bound on the decoding error probability which utilizes the sphere-packing bounding technique. This bound is valid for symmetric memoryless channels with a finite input alphabet, and is referred to as an improved sphere-packing (ISP) bound. We begin with some necessary definitions and basic properties of symmetric memoryless channels which are used in this section for the derivation of the ISP bound.

#### A. Symmetric Memoryless Channels

*Definition 3.1:* A bijective mapping $g : \mathcal{J} \to \mathcal{J}$ where $\mathcal{J} \subseteq \mathbb{R}^d$ is said to be *unitary* if for any integrable generalized function $f : \mathcal{J} \to \mathbb{R}$

$$\int_{\mathcal{J}} f(x) dx = \int_{\mathcal{J}} f(g(x)) dx \tag{23}$$

where by generalized function we mean a function which may contain a countable number of shifted Dirac delta functions. If the projection of $\mathcal{J}$ over some of the $d$ dimensions is countable, the integration over these dimensions is turned into a sum.

*Remark 3.1:* The following properties also hold:

1) If $g$ is a unitary mapping so is $g^{-1}$.
2) If $\mathcal{J}$ is a countable set, then $g : \mathcal{J} \to \mathcal{J}$ is unitary if and only if $g$ is bijective.
3) Let $\mathcal{J}$ be an open set and $g : \mathcal{J} \to \mathcal{J}$ be a bijective function. Denote

$$g(x_1, \ldots, x_d) \triangleq \big(g_1(x_1, \ldots, x_d), \ldots, g_d(x_1, \ldots, x_d)\big)$$

and assume that the partial derivatives $\frac{\partial g_i}{\partial x_j}$ exist for all $i, j \in \{1, 2, \ldots, d\}$. Then $g$ is unitary if and only if the Jacobian satisfies $|J(\mathbf{x})| = 1$ for all $\mathbf{x} \in \mathcal{J}$.

*Proof:* The first property follows from (23) and by defining $\widetilde{f}(x) \triangleq f\big(g^{-1}(x)\big)$; this gives

$$\int_{\mathcal{J}} f\big(g^{-1}(x)\big) dx = \int_{\mathcal{J}} \widetilde{f}(x) dx = \int_{\mathcal{J}} \widetilde{f}(g(x)) dx = \int_{\mathcal{J}} f\big((g^{-1} \circ g)(x)\big) dx = \int_{\mathcal{J}} f(x) dx .$$

The second property stems from the fact that for countable sets, the integral is turned into a sum, and the equality

$$\sum_{j \in \mathcal{J}} f(j) = \sum_{j \in \mathcal{J}} f\big(g(j)\big)$$

holds by changing the order of summation. Finally, the third property is proved by a transform of the integrator in the LHS of (23) from $\mathbf{x} = (x_1, \ldots, x_d)$ to $\big(g_1(x_1, \ldots, x_d), \ldots, g_d(x_1, \ldots, x_d)\big)$. ∎

We are now ready to define K-ary input symmetric channels. The symmetry properties of these channels are later exploited to improve the tightness of the sphere-packing bounding technique and derive the ISP lower bound on the average decoding error probability of block codes transmitted over these channels.

*Definition 3.2 (Symmetric Memoryless Channels):* A memoryless channel with input alphabet $\mathcal{K} = \{0, 1, \ldots, K-1\}$, output alphabet $\mathcal{J} \subseteq \mathbb{R}^d$ (where $K, d \in \mathbb{N}$) and transition probability (or density if $\mathcal{J}$ non-countable) $P(\cdot|\cdot)$ is said to be *symmetric* if there exists a set of unitary (bijective) mappings $\{g_k\}_{k=0}^{K-1}$ where $g_k : \mathcal{J} \to \mathcal{J}$ for all $k \in \mathcal{K}$ such that

$$\forall \mathbf{y} \in \mathcal{J}, \, k \in \mathcal{K} \quad P(\mathbf{y}|0) = P\big(g_k(\mathbf{y})|k\big) \tag{24}$$

and

$$\forall k_1, k_2 \in \mathcal{K} \quad g_{k_1}^{-1} \circ g_{k_2} = g_{(k_2 - k_1) \bmod K} . \tag{25}$$

*Remark 3.2:* From (24), the mapping $g_0$ is the identity mapping. Assigning $k_1 = k$ and $k_2 = 0$ in (25) gives

$$\forall k \in \mathcal{K} \quad g_k^{-1} = g_{(-k) \bmod K} = g_{K-k} . \tag{26}$$

The class of symmetric memoryless channels, as given in Definition 3.2, is quite large. In particular, it contains the class of MBIOS channels. To show this, we employ the following proposition (see [21, Section 4.1.4]):

*Proposition 3.1:* An arbitrary MBIOS channel can be equivalently represented as a binary symmetric channel (BSC) whose crossover probability is i.i.d., independent from the channel input, and observed by the receiver. The crossover probability is given by $p = \frac{1}{1+e^{|L|}}$ where $L$ denotes the random variable of the log-likelihood ratio at the channel output.

We now apply Proposition 3.1 to show that any MBIOS channel is a symmetric memoryless channel, according to Definition 3.2.

*Corollary 3.1:* An arbitrary MBIOS channel, can be equivalently represented as a symmetric memoryless channel.

*Proof:* Let us consider an MBIOS channel $\mathfrak{C}$. Applying Proposition 3.1, $\mathfrak{C}$ can be equivalently represented by a channel $\mathfrak{C}'$ whose output alphabet is $\mathcal{J} = \{0, 1\} \times [0, 1]$; here, the first term of the output refers to the BSC output and the second term is the associated crossover probability. We now show that this equivalent channel is a symmetric memoryless channel. To this end, it suffices to find a unitary mapping $g_1 : \mathcal{J} \to \mathcal{J}$ such that

$$\forall \mathbf{y} \in \mathcal{J} \quad P(\mathbf{y}|0) = P\big(g_1(\mathbf{y})|1\big) \tag{27}$$

and $g_1^{-1} = g_1$ (i.e., $g_1$ is equal to its inverse).

For the channel $\mathfrak{C}'$, the conditional probability distribution (or density) function of the output $\mathbf{y} = (m, p)$ (where $m \in \{0, 1\}$ and $p \in [0, 1]$) given that $i \in \{0, 1\}$ is transmitted, is given by

$$P(\mathbf{y}|i) = \begin{cases} \widetilde{P}(p) \cdot (1-p) & \text{if } i = m \\ \widetilde{P}(p) \cdot p & \text{if } i = \overline{m} \end{cases} \tag{28}$$

where $\widetilde{P}$ is a distribution (or density) over $[0, 1]$ and $\overline{m}$ designates the logical not of $m$. From (28), we get that the mapping $g_1(m, p) = \big(\overline{m}, p\big)$ satisfies (27). Additionally, $g_1^{-1} = g_1$ since $\overline{\overline{m}} = m$. Therefore, the proof is completed by showing that $g_1$ is a unitary mapping. For any (generalized) function $f : \mathcal{J} \to \mathbb{R}$ we have

$$\begin{aligned} \int_{\mathcal{J}} f(\mathbf{x}) d\mathbf{x} &\triangleq \sum_{m=0}^{1} \int_0^1 f(m, p) dp \\ &= \sum_{m=0}^{1} \int_0^1 f(\bar{m}, p) dp \\ &= \int_{\mathcal{J}} f\big(g_1(\mathbf{x})\big) d\mathbf{x} \end{aligned}$$

where the second equality holds by changing the order of summation; hence $g_1$ is a unitary function. ∎

*Remark 3.3:* Proposition 3.1 forms a special case of a proposition given in [36, Appendix I]. Using the proposition in [36, Appendix I], which refers to M-ary input channels, it can be shown in a similar way that all M-ary input symmetric output channels, as defined in [36], can be equivalently represented as symmetric memoryless channels. M-ary PSK modulated signals transmitted over the AWGN channel and coherently detected at the receiver form another example of a symmetric memoryless channel. In this case, $\mathcal{J}$ is defined to be $\mathbb{R}^2$ and $g_k$ is a clockwise rotation by $\frac{2\pi k}{M}$ where the determinant of the Jacobian is equal in absolute value to 1.

### B. Derivation of an Improved Sphere-Packing Bound for Symmetric Memoryless Channels

In this section, we derive an improved sphere-packing lower bound on the decoding error probability of block codes transmitted over symmetric memoryless channels. To keep the notation simple, we derive the bound under the assumption that the communication takes place over a symmetric DMC. However, the derivation of the bound is justified later for the general class of symmetric memoryless channels with discrete or continuous output alphabets. Some remarks are given at the end of the derivation.

Though there is a certain parallelism to the derivation of the SP67 bound in [25, Part 1], our analysis for symmetric memoryless channels deviates considerably from the derivation of this classical bound. The improvements suggested in [35] are also incorporated into the derivation of the bound. We show that for symmetric memoryless channels, the derivation of the sphere-packing bound can be modified so that the intermediate step of bounding the maximal error probability for fixed composition codes can be skipped, and one can directly consider the *average* error probability of an *arbitrary* block code. To this end, the first step of the derivation in [25] (see Theorem 2.1 here) is modified so that instead of bounding the error probability when a single pair of probability assignments is considered, we consider the average error probability over $M$ pairs of probability assignments.

*1) Average Decoding Error Probability for $M$ Pairs of Probability Assignments:* We start the analysis by considering the average decoding error probability over $M$ pairs of probability assignments, denoted $\{P_1^m, P_2^m\}_{m=1}^M$, where it is assumed that the index $m$ of the pair is chosen uniformly at random from the set $\{1, \ldots, M\}$ and is known to the decoder. Denote the observation by $\mathbf{y}$ and the observation space by $\mathcal{Y}$. For simplicity, we assume that $\mathcal{Y}$ is a finite set. Following the notation in [25], we define the LLR for the $m^{\text{th}}$ pair of probability assignments as

$$D^m(\mathbf{y}) \triangleq \ln\left(\frac{P_2^m(\mathbf{y})}{P_1^m(\mathbf{y})}\right) \tag{29}$$

and the probability distribution

$$Q_s^m(\mathbf{y}) \triangleq \frac{P_1^m(\mathbf{y})^{1-s}\, P_2^m(\mathbf{y})^s}{\sum_{\mathbf{y}'} P_1^m(\mathbf{y}')^{1-s}\, P_2^m(\mathbf{y}')^s}, \quad 0 \le s \le 1. \tag{30}$$

For the $m^{\text{th}}$ pair, we also define the function $\mu^m$ as

$$\mu^m(s) \triangleq \ln\left(\sum_{\mathbf{y}'} P_1^m(\mathbf{y}')^{1-s} P_2^m(\mathbf{y}')^s\right), \quad 0 \le s \le 1. \tag{31}$$

Let us assume that $\mu^m$ and its first and second derivatives w.r.t. $s$ are independent of the value of $m$, and therefore we can define $\mu \triangleq \mu^1 = \mu^2 = \ldots = \mu^M$.

*Remark 3.4:* Note that in this setting, the requirement that $\mu^m$ is independent of $m$ inherently yields that all its derivatives are also independent of $m$. However, in the continuation, we will let $P_2^m$ be a function of $s$ and differentiate $\mu^m$ w.r.t. $s$ while holding $P_2^m$ fixed. In this setting, we will show that for the specific selection of $P_1^m$ and $P_2^m$ which are used to derive the new lower bound on the average block error probability, if the communication takes place over a symmetric memoryless channel then $\mu^m$ and its first two derivatives w.r.t. $s$ are independent of $m$. Also note that the fact that $\mu^m$ is independent of $m$ does not imply that $P_k^m$ is independent of $m$.

Based on the assumption above, it can be easily verified (in parallel to (9)–(12)) that for all $m \in \{1, \ldots, M\}$

$$\mu'(s) = \left(\mu^m\right)'(s) = \mathbb{E}_{Q_s^m}\left(D^m(\mathbf{y})\right) \tag{32}$$

$$\mu''(s) = \left(\mu^m\right)''(s) = \operatorname{Var}_{Q_s^m}\left(D^m(\mathbf{y})\right) \tag{33}$$

$$P_1^m(\mathbf{y}) = \exp\left(\mu(s) - sD^m(\mathbf{y})\right) Q_s^m(\mathbf{y}) \tag{34}$$

$$P_2^m(\mathbf{y}) = \exp\left(\mu(s) + (1-s)D^m(\mathbf{y})\right) Q_s^m(\mathbf{y}) \tag{35}$$

where $\mathbb{E}_Q$ and $\operatorname{Var}_Q$ stand, respectively, for the statistical expectation and variance w.r.t. a probability distribution $Q$. For the $m^{\text{th}}$ code book, we define the set of typical output vectors as

$$\mathcal{Y}_s^{m,x} \triangleq \left\{\mathbf{y} \in \mathcal{Y} : |D^m(\mathbf{y}) - \mu'(s)| \le x\sqrt{2\mu''(s)}\right\}, \quad x > 0. \tag{36}$$

In the original derivation of the SP67 bound in [25] (see (13) here), the parameter $x$ was set to one; similarly to [35], this parameter is introduced in (36) for enabling to tighten the bound for finite-length block codes. However, in both [25] and [35], only one pair of probability assignments was considered. By applying Chebychev's inequality to (36), and relying on the equalities in (32) and (33), we get that for all $m \in \{1, \ldots, M\}$

$$\sum_{\mathbf{y} \in \mathcal{Y}_s^{m,x}} Q_s^m(\mathbf{y}) > 1 - \frac{1}{2x^2} \tag{37}$$

where this result is meaningful only for $x > \frac{\sqrt{2}}{2}$.

Let $\mathcal{Y}_1^m$ and $\mathcal{Y}_2^m$ be the decoding regions of $P_1^m$ and $P_2^m$, respectively. Since the index $m$ is known to the decoder, $P_1^m$ is decoded only against $P_2^m$; hence, $\mathcal{Y}_1^m$ and $\mathcal{Y}_2^m$ form a partition of the observation space $\mathcal{Y}$. We now derive a lower bound on the conditional error probability given that the correct hypothesis is the first probability assignment and the $m^{\text{th}}$ pair was selected. Similarly to (14), we get the following lower bound from (34) and (36):

$$P_{e,1}^m \ge \exp\left(\mu(s) - s\mu'(s) - s\,x\,\sqrt{2\mu''(s)}\right) \sum_{\mathbf{y} \in \mathcal{Y}_2^m \bigcap \mathcal{Y}_s^{m,x}} Q_s^m(\mathbf{y}). \tag{38}$$

Following the same steps w.r.t. the conditional error probability of $P_2^m$ and applying (35), gives

$$P_{\mathrm{e},1}^m \geq \exp\Big(\mu(s) + (1-s)\mu'(s) - (1-s)\, x\, \sqrt{2\mu''(s)}\Big) \sum_{\mathbf{y} \in \mathcal{Y}_1^m \bigcap \mathcal{Y}_s^{m,x}} Q_s^m(\mathbf{y}). \tag{39}$$

Averaging (38) and (39) over $m$ gives that for all $s \in (0,1)$

$$P_{\mathrm{e},1}^{\mathrm{avg}} \triangleq \frac{1}{M}\sum_{m=1}^M P_{\mathrm{e},1}^m \geq \exp\Big(\mu(s) - s\mu'(s) - s\, x\, \sqrt{2\mu''(s)}\Big) \frac{1}{M}\sum_{m=1}^M \sum_{\mathbf{y} \in \mathcal{Y}_1^m \bigcap \mathcal{Y}_s^{m,x}} Q_s^m(\mathbf{y}) \tag{40}$$

and

$$P_{\mathrm{e},2}^{\mathrm{avg}} \triangleq \frac{1}{M}\sum_{m=1}^M P_{\mathrm{e},2}^m \geq \exp\Big(\mu(s) + (1-s)\mu'(s) - (1-s)\, x\, \sqrt{2\mu''(s)}\Big) \frac{1}{M}\sum_{m=1}^M \sum_{\mathbf{y} \in \mathcal{Y}_1^m \bigcap \mathcal{Y}_s^{m,x}} Q_s^m(\mathbf{y}) \tag{41}$$

where $P_{\mathrm{e},1}^{\mathrm{avg}}$ and $P_{\mathrm{e},2}^{\mathrm{avg}}$ refer to the average error probabilities given that the first or second hypotheses, respectively, of a given pair are correct where this pair is chosen uniformly at random among the $M$ possible pairs of hypotheses. Since for all $m$, the sets $\mathcal{Y}_1^m$ and $\mathcal{Y}_2^m$ form a partition of the set of output vectors $\mathcal{Y}$, then

$$\frac{1}{M}\sum_{m=1}^M \sum_{\mathbf{y} \in \mathcal{Y}_1^m \bigcap \mathcal{Y}_s^{m,x}} Q_s^m(\mathbf{y}) + \frac{1}{M}\sum_{m=1}^M \sum_{\mathbf{y} \in \mathcal{Y}_2^m \bigcap \mathcal{Y}_s^{m,x}} Q_s^m(\mathbf{y}) = \frac{1}{M}\sum_{m=1}^M \sum_{\mathbf{y} \in \mathcal{Y}_s^{m,x}} Q_s^m(\mathbf{y}) > 1 - \frac{1}{2x^2}$$

where the last transition follows from (37) and is meaningful for $x > \frac{\sqrt{2}}{2}$. Hence, at least one of the terms in the LHS of the above equality is necessarily greater than $\frac{1}{2}\left(1 - \frac{1}{2x^2}\right)$. Combining this result with (40) and (41), we get that for every $s \in (0,1)$

$$P_{\mathrm{e},1}^{\mathrm{avg}} > \left(\frac{1}{2} - \frac{1}{4x^2}\right)\exp\Big(\mu(s) - s\mu'(s) - s\, x\, \sqrt{2\mu''(s)}\Big) \tag{42}$$

or

$$P_{\mathrm{e},2}^{\mathrm{avg}} > \left(\frac{1}{2} - \frac{1}{4x^2}\right)\exp\Big(\mu(s) + (1-s)\mu'(s) - (1-s)\, x\, \sqrt{2\mu''(s)}\Big). \tag{43}$$

The two inequalities above provide a lower bound on the average decoding error probability over $M$ pairs of probability assignments. We now turn to consider a block code which is transmitted over a symmetric DMC. Similarly to the derivation of the SP67 bound in [25], we use the lower bound derived in this section to relate the decoding error probability when a given codeword is transmitted to the size of the decision region associated with this codeword. However, the bound above allows us to directly consider the average block error probability; this is in contrast to the derivation in [25] which first considered the maximal block error probability of the code and then used an argument based on expurgating half of the bad codewords in order to obtain a lower bound on the average error probability of the expurgated code (where the code rate is asymptotically not affected as a result of this expurgation). Additionally, we show that when the transmission takes place over a memoryless symmetric channel, one can consider directly an arbitrary block code instead of starting the analysis by referring to fixed composition codes as in [25, Part 1] and [35].

*2) Lower Bound on the Decoding Error Probability of General Block Codes:* We now consider a block code $\mathcal{C}$ of length $N$ with $M$ codewords, denoted by $\{\mathbf{x}_m\}_{m=1}^M$; assume that the transmission takes place over a symmetric DMC with transition probabilities $P(j|k)$, where $k \in \mathcal{K} = \{0, \ldots, K-1\}$ and $j \in \mathcal{J} = \{0, \ldots, J-1\}$ designate the channel input and output alphabets, respectively. In this section, we derive a lower bound on the average block error probability of the code $\mathcal{C}$ under an arbitrary list decoder where the size of the list is limited to $L$. Let $f_N$ be a probability measure defined over the set of length-$N$ sequences of the channel output, and which can be factorized as in (16). We define $M$ pairs of probability measures $\{P_1^m, P_2^m\}$ by

$$P_1^m(\mathbf{y}) \triangleq \Pr(\mathbf{y}|\mathbf{x}_m), \quad P_2^m(\mathbf{y}) \triangleq f_N(\mathbf{y}), \quad m \in \{1, 2, \ldots, M\} \tag{44}$$

where $\mathbf{x}_m$ is the $m^{\mathrm{th}}$ codeword of the code $\mathcal{C}$. Combining (31) and (44), the function $\mu^m$ takes the form

$$\mu^m(s) = \ln\left(\sum_{\mathbf{y}} \Pr(\mathbf{y}|\mathbf{x}_m)^{1-s} f_N(\mathbf{y})^s\right), \quad 0 < s < 1. \tag{45}$$

Let us denote by $q_k^m$ the fraction of appearances of the letter $k$ in the codeword $\mathbf{x}_m$. By assumption, the communication channel is memoryless and the function $f_N$ is a probability measure which is factorized according to (16). Hence, for every $m \in \{1, 2, \ldots, M\}$, the function $\mu^m(s)$ in (45) is expressible in the form

$$\mu^m(s) = N \sum_{k=0}^{K-1} q_k^m \mu_k(s) \tag{46}$$

where

$$\mu_k(s) \triangleq \ln \left( \sum_{j=0}^{J-1} P(j|k)^{1-s} f(j)^s \right), \quad 0 < s < 1. \tag{47}$$

In order to apply the bound in (42) and (43), it is required that the function $\mu^m$ and its first and second derivatives w.r.t. $s$ are independent of the index $m$. From (46), it suffices to show that $\mu_k$ and its first and second derivatives are independent of the input symbol $k$. To this end, for every $s \in (0, 1)$, we choose the function $f$ to be $f_s$, as given in [25, Eqs. (4.18)–(4.20)]. Namely, for $0 < s < 1$, let $\mathbf{q}_s = \{q_{0,s}, \ldots, q_{K-1,s}\}$ satisfy the inequalities

$$\sum_{j=0}^{J-1} P(j|k)^{1-s} \alpha_{j,s}^{\frac{s}{1-s}} \geq \sum_{j=0}^{J-1} \alpha_{j,s}^{\frac{1}{1-s}} \; ; \quad \forall k \tag{48}$$

where

$$\alpha_{j,s} \triangleq \sum_{k'=0}^{K-1} q_{k',s} P(j|k')^{1-s}. \tag{49}$$

The function $f = f_s$ is given by

$$f_s(j) = \frac{\alpha_{j,s}^{\frac{1}{1-s}}}{\sum_{j'=0}^{J-1} \alpha_{j',s}^{\frac{1}{1-s}}}, \quad j \in \{0, \ldots, J-1\}. \tag{50}$$

Note that the input distribution $\mathbf{q}_s$ is *independent of the code* $\mathcal{C}$, as it only depends on the channel statistics. It should also be noted that since the bound in (42) and (43) holds for every $s \in (0, 1)$, $P_1^m$ and $P_2^m$ are in general allowed to depend on $s$. However, the differentiation of the function $\mu^m$ w.r.t. $s$ is performed while holding $P_1^m$ and $P_2^m$ fixed. The following lemma shows that for symmetric channels, the function $f_s$ in (50) yields that $\mu_k$ and its first and second derivatives w.r.t. $s$ (while holding $f_s$ fixed) are independent of the input symbol $k$.

*Lemma 3.1:* Let $P(\cdot|\cdot)$ designate the transition probability function of a symmetric DMC with input alphabet $\mathcal{K} = \{0, \ldots, K-1\}$ and output alphabet $\mathcal{J} = \{0, \ldots, J-1\}$, and let $\mu_k$ be defined as in (47), where $f = f_s$ is given in (50). Then, the following properties hold for all $s \in (0, 1)$

$$\mu_0(s) = \mu_1(s) = \ldots = \mu_{K-1}(s) = -(1-s)E_0\left(\frac{s}{1-s}\right) \tag{51}$$

$$\mu_0'(s) = \mu_1'(s) = \ldots = \mu_{K-1}'(s) \tag{52}$$

$$\mu_0''(s) = \mu_1''(s) = \ldots = \mu_{K-1}''(s) \tag{53}$$

where $E_0$ is introduced in (19) and the differentiation in (52) and (53) is performed w.r.t $s$ while holding $f_s$ fixed.

*Proof:* The proof of this lemma is quite technical and is given in Appendix A. ∎

*Remark 3.5:* Since the differentiation of the function $\mu_k$ w.r.t. $s$ is performed while holding $f = f_s$ fixed, then the independence of the function $\mu_k$ in the parameter $k$, as stated in (51), does not necessarily imply the independence of the first and second derivatives of $\mu_k$ as in (52) and (53); in order to prove this lemma, we rely on the symmetry of the memoryless channel. The function $\mu_0$ in (4) and its derivatives are calculated in Appendix B for some symmetric memoryless channels, and these results are later used for the numerical calculations of the sphere-packing bounds in Section V.

By (46) and Lemma 3.1 we get that the function $\mu^m$ and its first and second derivatives w.r.t. $s$ are independent of the index $m$ (where this property also follows since $\sum_{k=0}^{K-1} q_k^m = 1$, irrespectively of $m$). Hence, the lower bound derived in Section III-B.1 can be applied.

Let $\mathcal{Y}_m$ be the decision region of the codeword $\mathbf{x}_m$. By associating $\mathcal{Y}_m$ and $\mathcal{Y}_m^c$ with the two decision regions for the probability measures $P_1^m$ and $P_2^m$, respectively, we get

$$P_{\mathrm{e},1}^m = \sum_{\mathbf{y} \in \mathcal{Y}_m^c} P_1^m(\mathbf{y}) = \sum_{\mathbf{y} \in \mathcal{Y}_m^c} \Pr(\mathbf{y}|\mathbf{x}_m) \triangleq P_{\mathrm{e},m}$$

and

$$P_{\mathrm{e},2}^m = \sum_{\mathbf{y} \in \mathcal{Y}_m} P_2^m(\mathbf{y}) = \sum_{\mathbf{y} \in \mathcal{Y}_m} f_N(\mathbf{y}) = F(\mathcal{Y}_m)$$

where $P_{\mathrm{e},m}$ is the decoding error probability of the code $\mathcal{C}$ when the codeword $\mathbf{x}_m$ is transmitted, and $F(\mathcal{Y}_m)$ is a measure for the size of the decoding region $\mathcal{Y}_m$ as defined in (17). Substituting the two equalities above in (42) and (43) gives that for all $s \in (0,1)$

$$\frac{1}{M} \sum_{m=1}^{M} P_{\mathrm{e},m} = P_{\mathrm{e},1}^{\mathrm{avg}} > \left(\frac{1}{2} - \frac{1}{4x^2}\right) \exp\Big(\mu(s) - s\mu'(s) - s\,x\,\sqrt{2\mu''(s)}\Big) \tag{54}$$

or

$$\frac{1}{M} \sum_{m=1}^{M} F_s(\mathcal{Y}_m) = P_{\mathrm{e},2}^{\mathrm{avg}} > \left(\frac{1}{2} - \frac{1}{4x^2}\right) \exp\Big(\mu(s) + (1-s)\mu'(s) - (1-s)\,x\,\sqrt{2\mu''(s)}\Big) \tag{55}$$

where $x > \frac{\sqrt{2}}{2}$ and $F_s(\mathcal{Y}_m) \triangleq \sum_{\mathbf{y} \in \mathcal{Y}_m} f_{N,s}(\mathbf{y})$. Similarly to [25], we relate $\sum_{m=1}^{M} F_s(\mathcal{Y}_m)$ to the number of codewords $M$ and to the size of the decoding list which is limited to $L$. First, for all $0 \le s \le 1$

$$\sum_{m=1}^{M} F_s(\mathcal{Y}_m) = \sum_{m=1}^{M} \sum_{\mathbf{y} \in \mathcal{Y}_m} f_{N,s}(\mathbf{y}) \le L$$

where the last inequality holds since each $\mathbf{y} \in \mathcal{J}^N$ is included in at most $L$ subsets $\{\mathcal{Y}_m\}_{m=1}^{M}$ and also $\sum_{\mathbf{y}} f_{N,s}(\mathbf{y}) = 1$. Hence, the LHS of (55) is upper bounded by $\frac{L}{M}$ for all $0 \le s \le 1$. Additionally, the LHS of (54) is equal by definition to the average block error probability $P_{\mathrm{e}}$ of the code $\mathcal{C}$. Therefore, (54) and (55) can be rewritten as

$$P_{\mathrm{e}} > \left(\frac{1}{2} - \frac{1}{4x^2}\right) \exp\Big(\mu(s) - s\mu'(s) - s\,x\,\sqrt{2\mu''(s)}\Big) \tag{56}$$

or

$$\frac{L}{M} > \left(\frac{1}{2} - \frac{1}{4x^2}\right) \exp\Big(\mu(s) + (1-s)\mu'(s) - (1-s)\,x\,\sqrt{2\mu''(s)}\Big). \tag{57}$$

Applying (46) and Lemma 3.1 to (56) and (57) gives that for all $s \in (0,1)$

$$P_{\mathrm{e}} > \left(\frac{1}{2} - \frac{1}{4x^2}\right) \exp\left\{ N \left(\mu_0(s,f_s) - s\mu_0'(s,f_s) - s\,x\,\sqrt{\frac{2\mu_0''(s,f_s)}{N}}\right)\right\} \tag{58}$$

or

$$\frac{L}{M} > \left(\frac{1}{2} - \frac{1}{4x^2}\right) \exp\left\{ N \left(\mu_0(s,f_s) + (1-s)\mu_0'(s,f_s) - (1-s)\,x\,\sqrt{\frac{2\mu_0''(s,f_s)}{N}}\right)\right\}. \tag{59}$$

A lower bound on the average block error probability can be obtained from (58) by substituting any value of $s \in (0,1)$ for which the inequality in (59) does not hold. In particular we choose a value $s = s_x$ such that the inequality in (59) is replaced by equality, i.e.,

$$\begin{aligned}
\frac{L}{M} &= \exp(-NR) \\
&= \left(\frac{1}{2} - \frac{1}{4x^2}\right) \exp\left\{ N\Big(\mu_0(s_x,f_{s_x}) + (1-s_x)\,\mu_0'(s_x,f_{s_x}) - (1-s_x)\,x\,\sqrt{\frac{2\mu_0''(s_x,f_{s_x})}{N}}\Big)\right\}
\end{aligned} \tag{60}$$

where $R \triangleq \frac{\ln\left(\frac{M}{L}\right)}{N}$ designates the code rate in nats per channel use. Note that the existence of a solution $s = s_x$ to (60) can be demonstrated in a similar way to the arguments in [25, Eqs. (4.28)–(4.35)] for the non-trivial case where the sphere-packing bound does not reduce to the trivial inequality $P_e \geq 0$. This particular value of $s$ is chosen since for a large enough value of $N$, the RHS of (58) is monotonically decreasing while the RHS of (59) is monotonically increasing for $s \in (0, 1)$; thus, this choice is optimal for large enough $N$. The choice of $s = s_x$ also allows to get a simpler representation of the bound on the average block error probability. Rearranging (60) gives

$$\mu_0'(s_x, f_{s_x}) = -\frac{1}{1 - s_x} \left[ R + \mu_0(s_x, f_{s_x}) + \frac{1}{N} \, \ln\left(\frac{1}{2} - \frac{1}{4x^2}\right) \right] + x \sqrt{\frac{2\mu_0''(s_x, f_{s_x})}{N}}.$$

Substituting $s = s_x$ and the last equality into (58) yields that

$$
\begin{aligned}
P_e \; > \; \exp\Bigg\{ N \, \bigg( &\frac{\mu_0(s_x, f_{s_x})}{1 - s_x} + \frac{s_x}{1 - s_x} \left( R + \frac{1}{N} \ln\left(\frac{1}{2} - \frac{1}{4x^2}\right)\right) \\
&- s_x \, x \, \sqrt{\frac{8\mu_0''(s_x, f_{s_x})}{N}} + \frac{1}{N} \ln\left(\frac{1}{2} - \frac{1}{4x^2}\right) \bigg) \Bigg\}.
\end{aligned}
$$

By applying (51) and defining $\rho_x \triangleq \frac{s_x}{1 - s_x}$ we get

$$
\begin{aligned}
P_e \; > \; \exp\Bigg\{ -N \, \bigg( &E_0(\rho_x) - \rho_x \left[ R - \frac{\ln 4}{N} + \frac{\ln\left(2 - \frac{1}{x^2}\right)}{N} \right] \\
&+ s_x \, x \, \sqrt{\frac{8\mu_0''(s_x, f_{s_x})}{N}} + \frac{\ln 4}{N} - \frac{\ln\left(2 - \frac{1}{x^2}\right)}{N} \bigg) \Bigg\}.
\end{aligned}
$$

Note that the above lower bound on the average decoding error probability holds for an arbitrary block code of length $N$ and rate $R$. The selection of $\rho_x$ is similar to [35]. Finally, we optimize over the parameter $x \in (\frac{\sqrt{2}}{2}, \infty)$ in order to get the tightest lowest bound of this form.

The derivation above only relies on the fact that the channel is memoryless and symmetric, but does not rely on the fact that the output alphabet is discrete. As mentioned in Section II-B, the original derivation of the SP67 bound in [25] relies on the fact that the input and output alphabets are finite in order to upper bound $\mu''(s)$ by $\left(\frac{1}{s} \ln\left(\frac{e}{\sqrt{P_{\min}}}\right)\right)^2$ where $P_{\min}$ designates the smallest non-zero transition probability of the channel. This requirement was relaxed in [35] to the requirement that only the input alphabet is finite; to this end, the second derivative of the function $\mu$ is calculated, thus the above upper bound on this second derivative is replaced by its exact value. The validity of the derivation for symmetric continuous-output channels is provided in the continuation (see Remark 3.9). This leads to the following theorem, which provides an improved sphere-packing lower bound on the decoding error probability of block codes transmitted over symmetric memoryless channels.

*Theorem 3.1 (An Improved Sphere-Packing (ISP) Bound for Symmetric Memoryless Channels):* Let $\mathcal{C}$ be an arbitrary block code consisting of $M$ codewords, each of length $N$. Assume that $\mathcal{C}$ is transmitted over a memoryless symmetric channel which is specified by the transition probabilities (or densities) $P(j|k)$ where $k \in \mathcal{K} = \{0, \ldots, K - 1\}$ and $j \in \mathcal{J} \subseteq \mathbb{R}^d$ designate the channel input and output alphabets, respectively. Assume an arbitrary list decoder where the size of the list is limited to $L$. The *average decoding error probability* satisfies

$$P_e(N, M, L) \geq \exp\left\{ -N \widetilde{E}_{sp}(R, N) \right\}$$

where

$$\widetilde{E}_{sp}(R, N) \triangleq \sup_{x > \frac{\sqrt{2}}{2}} \left\{ E_0(\rho_x) - \rho_x \left( R - O_1\left(\frac{1}{N}, x\right) \right) + O_2\left(\frac{1}{\sqrt{N}}, x, \rho_x\right) \right\} \tag{61}$$

the function $E_0$ is introduced in (19), $R = \frac{1}{N} \, \ln\left(\frac{M}{L}\right)$, and

$$
\begin{aligned}
O_1\left(\frac{1}{N}, x\right) &\triangleq \frac{\ln 4}{N} - \frac{\ln\left(2 - \frac{1}{x^2}\right)}{N} \tag{62} \\
O_2\left(\frac{1}{\sqrt{N}}, x, \rho\right) &\triangleq s(\rho) \, x \, \sqrt{\frac{8}{N} \mu_0''\big(s(\rho), f_{s(\rho)}\big)} + \frac{\ln 4}{N} - \frac{\ln\left(2 - \frac{1}{x^2}\right)}{N}. \tag{63}
\end{aligned}
$$

Here, $s(\rho) \triangleq \frac{\rho}{1+\rho}$, the non-negative parameter $\rho = \rho_x$ in the RHS of (61) is determined by solving the equation

$$R - O_1\left(\frac{1}{N}, x\right) = -\mu_0\big(s(\rho), f_{s(\rho)}\big) - \big(1 - s(\rho)\big)\mu_0'\big(s(\rho), f_{s(\rho)}\big) + \big(1 - s(\rho)\big)\, x\, \sqrt{\frac{2\mu_0''\big(s(\rho), f_{s(\rho)}\big)}{N}} \qquad (64)$$

and the functions $\mu_0(s, f)$ and $f_s$ are defined in (47) and (50), respectively.

*Remark 3.6:* The requirement that the communication channel is symmetric is crucial to the derivation of the ISP bound. One of the new concepts introduced here is the use of the channel symmetry to show that the function $\mu^m$ and its first and second derivatives w.r.t. $s$ are independent of the codeword composition. This enables to tighten the VF bound in [35] by skipping the intermediate step which is related to fixed composition codes. Another new concept is a direct consideration of the *average decoding error probability* of the code rather than considering the maximal block error probability and expurgating the code. This is due to the consideration of $M$ pairs of probability distributions in the first step of the derivation. Note that the bound on the average block error probability of $M$ probability assignment pairs requires that $\mu^m$ and its first and second derivatives are independent of the index $m$; this property holds due to the symmetry of the memoryless communication channel.

*Remark 3.7:* In light of the previous remark where we do not need to consider the block error probability of fixed composition codes as an intermediate step, the ISP bound differs from the VF bound [35] (see Theorem 2.4) in the sense that the term $\frac{\log\binom{N+K-1}{K-1}}{N}$ is removed from $O_1(\frac{\ln N}{N}, x)$ (see (22)). Therefore, the shift in the rate of the error exponent of the ISP bound behaves asymptotically like $O_1\left(\frac{1}{N}\right)$ instead of $O_1\left(\frac{\ln N}{N}\right)$ (see (21), (22) and (62)). Additionally, the derivation of the VF bound requires expurgation of the code to transform a lower bound on the maximal block error probability to a lower bound on the average block error probability. These differences indicate a tightening of the pre-exponent of the ISP bound (as compared to the SP67 and VF bounds) which is expected to be especially pronounced for codes of small to moderate block lengths and also when the size of the channel input alphabet is large (as will be verified in Section V).

*Remark 3.8:* The rate loss as a result of the expurgation of the code by removing half of the codewords with the largest error probability was ignored in [35]. The term $\frac{\ln 4}{N}$, as it appears in the term $O_1(\frac{\ln N}{N}, x)$ of [35, Theorem 7], should be therefore replaced by $\frac{\ln 8}{N}$ (see (62)).

*Remark 3.9:* The ISP bound is also applicable to symmetric channels with continuous output. When the ISP bound is applied to a memoryless symmetric channel with a continuous-output alphabet, the transition probability is replaced by a transition density function and the sums over the output alphabet are replaced by integrals. Note that these densities may include Dirac delta functions which appear at the points where the corresponding input distribution or the transition density function of the channel are discontinuous. Additionally, as explained in Appendix A, the statement in Lemma 3.1 holds for general symmetric memoryless channels.

## IV. The 1959 Sphere-Packing Bound of Shannon and Improved Algorithms for Its Calculation

The 1959 sphere-packing (SP59) bound derived by Shannon [24] provides a lower bound on the decoding error probability of an arbitrary block code whose transmission takes place over an AWGN channel. We begin this section by introducing the SP59 bound in its original form, along with asymptotic approximations derived by Shannon [24] which facilitate the estimation of the bound for large block lengths. We then review a theorem, introduced by Valembois and Fossorier [35], presenting a set of recursive equations which simplify the calculation of this bound. Both the original formula for the SP59 bound in [24] and the recursive method in [35] perform the calculations in the probability domain; this leads to various numerical difficulties of over and under flows when calculating the exact value of the bound for codes of block lengths of $N = 1000$ or more. In this section, we present an alternative approach which facilitates the calculation of the SP59 bound in the logarithmic domain. This eliminates the possibility of numerical problems in the calculation of the SP59 bound, regardless of the block length.

### A. The 1959 Sphere-Packing Bound and Asymptotic Approximations

Consider a block code $\mathcal{C}$ of length $N$ and rate $R$ nats per channel use per dimension. It is assumed that all the codewords are mapped to signals with equal energy (e.g., PSK modulation); hence, all the signals representing codewords lie on an $N$-dimensional sphere centered at the origin, but finer details of the modulation used are not taken into account in the derivation of the bound. This assumption implies that every Voronoi cell (i.e., the convex region containing all the points which are closer to the considered signal than to any other code signal) is a polyhedric cone which is limited by at most $\exp(NR) - 1$ hyper planes intersecting at the origin. As a measure of volume, Shannon introduced the solid angle of a cone which is defined to be the area of the sphere of unit radius cut out by the cone. Since the Voronoi cells partition the space $\mathbb{R}^N$, then the sum of their solid angles is equal to the area of an $N$-dimensional sphere of unit radius. The derivation of the SP59 bound relies on two main observations:

- Among the cones of a given solid angle, the lowest probability of error is obtained by the circular cone whose axis connect the code signal with the origin.
- In order to minimize the average decoding error probability, it is best to share the total solid angle equally among the $\exp(NR)$ Voronoi regions.

As a corollary of these two observations, it follows that the average block error probability cannot be below the one which corresponds to the case where all the Voronoi regions are circular cones centered around the code signals with a common solid angle which is equal to a fraction of $\exp(-NR)$ of the solid angle of $\mathbb{R}^N$. The solid angle of a circular cone is given by the following lemma.

*Lemma 4.1 (Solid Angle of a Circular Cone [24]):* The solid angle of a circular cone of half angle $\theta$ in $\mathbb{R}^N$ is given by

$$\Omega_N(\theta) = \frac{2\pi^{\frac{N-1}{2}}}{\Gamma(\frac{N-1}{2})} \int_0^\theta (\sin\phi)^{N-2} \, d\phi \ .$$

In particular, the solid angle of $\mathbb{R}^N$ is given by

$$\Omega_N(\pi) = \frac{2\pi^{\frac{N}{2}}}{\Gamma(\frac{N}{2})} \ .$$

*Theorem 4.1 (The 1959 Sphere-Packing (SP59) Bound [24]):* Assume that the transmission of an arbitrary block code of length $N$ and rate $R$ (in units of nats per channel use per dimension) takes place over an AWGN channel with noise spectral density $\frac{N_0}{2}$. Then, under ML decoding, the block error probability is lower bounded by

$$P_e(\text{ML}) > P_{\text{SPB}}(N, \theta, A) \ , \quad A \triangleq \sqrt{\frac{2E_s}{N_0}} \tag{65}$$

where $E_s$ is the average energy per symbol, $\theta \in [0, \pi]$ satisfies the inequality $\exp(-NR) \leq \frac{\Omega_N(\theta)}{\Omega_N(\pi)}$,

$$P_{\text{SPB}}(N, \theta, A) \triangleq \frac{(N-1)e^{-\frac{NA^2}{2}}}{\sqrt{2\pi}} \int_\theta^{\frac{\pi}{2}} (\sin\phi)^{N-2} \, f_N(\sqrt{N}A\cos\phi) \, d\phi + Q(\sqrt{N}A) \tag{66}$$

and

$$f_N(x) \triangleq \frac{1}{2^{\frac{N-1}{2}}\Gamma(\frac{N+1}{2})} \int_0^\infty z^{N-1} \exp\left(-\frac{z^2}{2} + zx\right) dz \ , \quad \forall \, x \in \mathbb{R}, \ N \in \mathbb{N}. \tag{67}$$

By assumption, the transmitted signal is represented by a point which lies on the $N$-dimensional sphere of radius $\sqrt{NE_s}$ and which is centered at the origin, and the Gaussian noise is additive. The value $P_{\text{SPB}}(N, \theta, A)$ in the RHS of (65) designates the probability that the received vector falls outside the $N$-dimensional circular cone of half angle $\theta$ whose main axis passes through the origin and the signal point which is represented by the transmitted signal. Hence, this function is monotonically decreasing in $\theta$. The tightest lower bound on the decoding error probability is therefore achieved for $\theta_1(N, R)$ which satisfies

$$\frac{\Omega_N(\theta_1(N,R))}{\Omega_N(\pi)} = \exp(-NR). \tag{68}$$

The calculation of $\theta_1(N, R)$ can become quite tedious. In order to simplify the calculation of the SP59 bound, Shannon provided in [24] asymptotically tight upper and lower bounds on the ratio $\frac{\Omega_N(\theta)}{\Omega_N(\pi)}$.

*Lemma 4.2 (Bounds on the Solid Angle [24]):* The solid angle of a circular cone of half angle $\theta$ in the Euclidean space $\mathbb{R}^N$ satisfies the inequality

$$\frac{\Gamma(\frac{N}{2})(\sin\theta)^{N-1}}{2\Gamma(\frac{N+1}{2})\sqrt{\pi}\cos\theta} \left(1 - \frac{\tan^2\theta}{N}\right) \leq \frac{\Omega_N(\theta)}{\Omega_N(\pi)} \leq \frac{\Gamma(\frac{N}{2})(\sin\theta)^{N-1}}{2\Gamma(\frac{N+1}{2})\sqrt{\pi}\cos\theta} \ .$$

*Corollary 4.1 (SP59 Bound (Cont.)):* If $\theta^*$ satisfies the equation

$$\frac{\Gamma(\frac{N}{2})(\sin\theta^*)^{N-1}}{2\Gamma(\frac{N+1}{2})\sqrt{\pi}\cos\theta^*}\left(1-\frac{\tan^2\theta^*}{N}\right)=\exp(-NR) \tag{69}$$

then $\frac{\Omega_N(\theta^*)}{\Omega_N(\pi)}\geq\exp(-NR)$, and therefore

$$P_e(\text{ML})>P_{\text{SPB}}(N,\theta^*,A). \tag{70}$$

The use of $\theta^*$ instead of the optimal value $\theta_1(N,R)$ causes some loss in the tightness of the SP59 bound. However, due to the asymptotic tightness of the bounds on $\frac{\Omega_N(\theta)}{\Omega_N(\pi)}$, this loss vanishes as $N\to\infty$. In [35], it was numerically observed that this loss is marginal even for relatively small values of $NR$; it was observed that this loss is smaller then $0.01$ dB whenever the dimension of the code in bits is greater than 20, and it becomes smaller then $0.001$ dB when the dimension exceeds 60 bits.

For large block lengths, the calculation of the SP59 becomes practically difficult due to over and under flows in the floating-point operations. However, [24] presents some asymptotic formulas which give a good estimation of the bound for large enough block lengths. These approximations allow the calculation to be made in the logarithmic domain which eliminates the possibility of floating-point errors.

*Theorem 4.2:* [24]. Defining

$$G(\theta) \triangleq \frac{A\cos\theta+\sqrt{A^2\cos^2\theta+4}}{2}$$

$$E_L(\theta) \triangleq \frac{A^2-AG(\theta)\cos\theta-2\ln\big(G(\theta)\sin\theta\big)}{2}$$

then

$$P_{\text{SPB}}(N,\theta,A)\geq\frac{\sqrt{N-1}}{6N(A+1)}e^{\frac{-(A+1)^2+3}{2}}e^{-N\,E_L(\theta)}. \tag{71}$$

This lower bound is valid for any block length $N$. However, the ratio of the left and right terms in (71) stays bounded away from one for all $N$.

A rather accurate approximation of $P_{\text{SPB}}(N,\theta,A)$ was provided by Shannon in [24], but without a determined inequality. As a consequence, the following approximation is not a proven theoretical lower bound on the block error probability. For $N>1000$, however, its numerical values become almost identical to those of the exact bound, thus giving a useful estimation for the lower bound.

*Proposition 4.1:* [24]. Using the notation of Theorem 4.2, if $\theta>\cot^{-1}(A)$, then

$$P_{\text{SPB}}(N,\theta,A)\approx\frac{\alpha(\theta)e^{-NE_L(\theta)}}{\sqrt{N}}$$

where

$$\alpha(\theta)\triangleq\left(\sqrt{\pi\left(1+G(\theta)^2\right)}\,\sin\theta\,\big(AG(\theta)\sin^2\theta-\cos\theta\big)\right)^{-1}.$$

### B. A Recent Algorithm for Calculating the 1959 Sphere-Packing Bound

In [35, Section 2], Valembois and Fossorier review the SP59 bound and suggest a recursive algorithm to simplify its calculation. This algorithm is presented in the following theorem:

*Theorem 4.3 (Recursive Equations for Simplifying the Calculation of the SP59 Bound):* [35, Theorem 3]. The set of functions $\{f_N\}$ introduced in (67) can be expressed in the alternative form

$$f_N(x)=P_N(x)+Q_N(x)\exp(\frac{x^2}{2})\int_{-\infty}^{x}\exp(-\frac{t^2}{2})\,dt\,,\quad x\in\mathbb{R},\ N\in\mathbb{N} \tag{72}$$

where $P_N$ and $Q_N$ are two polynomials, determined by the same recursive equation for all $N\geq 5$

$$P_N(x)=\frac{2N-5+x^2}{N-1}\,P_{N-2}(x)-\frac{N-4}{N-1}\,P_{N-4}(x)\,,$$

$$Q_N(x)=\frac{2N-5+x^2}{N-1}\,Q_{N-2}(x)-\frac{N-4}{N-1}\,Q_{N-4}(x) \tag{73}$$

with the initial conditions

$$P_1(x) = 0, \quad P_2(x) = \sqrt{\frac{2}{\pi}}, \quad P_3(x) = \frac{x}{2}, \quad P_4(x) = \sqrt{\frac{2}{\pi}}\frac{2 + x^2}{3},$$

$$Q_1(x) = 1, \quad Q_2(x) = \sqrt{\frac{2}{\pi}}\,x, \quad Q_3(x) = \frac{1 + x^2}{2}, \quad Q_4(x) = \sqrt{\frac{2}{\pi}}\frac{3x + x^3}{3}.$$

By examining the recursive equations for $P_N$ and $Q_N$ in (73), it is noticed that the coefficients of the higher powers of $x$ vanish exponentially as $N$ increases. When performing the calculation using double-precision floating-point numbers, these coefficients cause underflows when $N$ is larger than several hundreds, and are replaced by zeros. Examining the expression for $P_{\mathrm{SPB}}(N, \theta, A)$ in (66), we observe that $f_N(x)$ (and therefore the polynomials $P_N(x)$ and $Q_N(x)$) is evaluated at $x \sim O(\sqrt{N})$. Hence, the replacement of the coefficients of the high powers of $x$ by zeros causes a considerable inaccuracy in the calculation of $P_{\mathrm{SPB}}$ in (66). To exemplify the effect of these underflows, we study the coefficients of $P_{750}(x)$ as calculated using double precision floating-point numbers. In this case, the coefficients of all the powers higher than 400 have caused underflows and have been replaced by zeros. The left plot of Figure 1 shows the coefficients of $P_{750}(x)$. Since $f_N(x)$ is evaluated at $x \sim O(\sqrt{N})$, one should examine the coefficients of $\tilde{P}_{750}(x) \triangleq P_{750}(\sqrt{750}\,x)$ which are plotted in the right plot of Figure 1. It can be seen that the dominant coefficients are those multiplying the powers of $x$ between 400 and 520 which, as mentioned above, have been replaced by zeros due to underflows. This demonstrates the inaccuracy due to underflows in the coefficients of the high powers. To avoid this loss of dominant coefficients, it is possible to modify the recursive equations (73) in order to calculate the polynomials $\tilde{P}_N(x) \triangleq P_N(\sqrt{N}\,x)$ and $\tilde{Q}_N(x) \triangleq Q(\sqrt{N}\,x)$. However, as observed in the right plot of Figure 1, these coefficients become extremely large and cause overflows when $N$ approaches 1000.

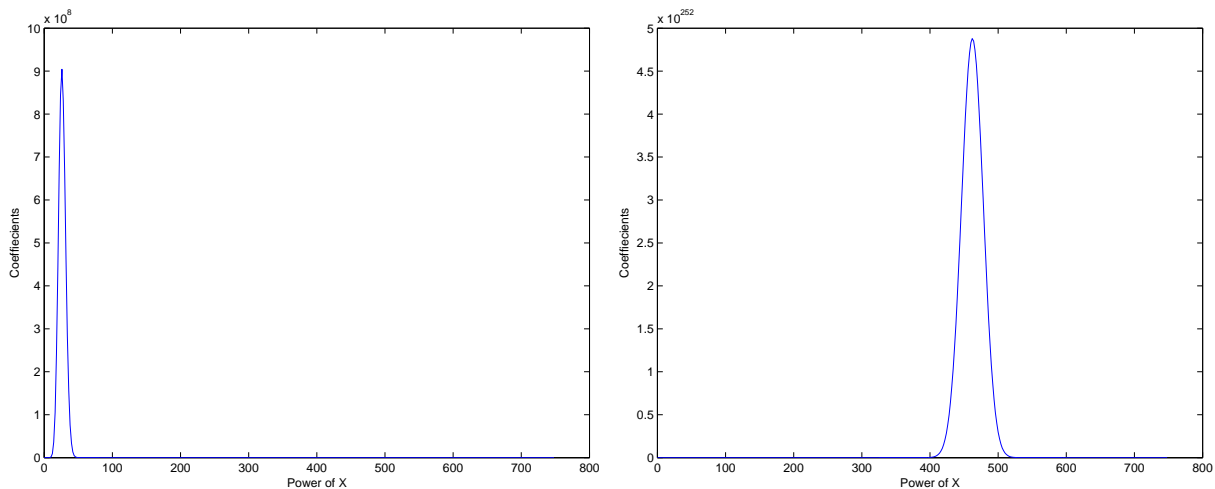

Fig. 1. Coefficients of the polynomials $P_{750}(x)$ (left plot) and $\tilde{P}_{750}(x) = P_{750}(\sqrt{750}\,x)$ (right plot). Since the polynomials are even, only the coefficients multiplying the even powers of $x$ have been plotted. It can be observed that in the right plot, the coefficients of powers of $x$ between 400 and 520 are dominant. These coefficients have caused underflows in the calculation of $P_{750}(x)$ in the left plot.

Considering the integrand in the RHS of (66) reveals another difficulty in calculating the SP59 bound for large values on $N$. For these values, the term $f_N(\sqrt{N}A\cos\phi)$ becomes very large and causes overflows, while the value of the term $(\sin\phi)^{N-2}$ becomes very small and causes underflows; this causes a "$0 \cdot \infty$" phenomenon when evaluating the integrand at the RHS of (66).

## C. A Log-Domain Approach for Computing the 1959 Sphere-Packing Bound

In this section, we present a method which enables the entire calculation of the integrand in the RHS of (66) in the log domain, thus circumventing the numerical over and under flows which become problematic in the calculation of the SP59 bound for large block lengths. We begin our derivation by representing the set of functions $\{f_N\}$ defined in (67) as sums of exponents.

*Proposition 4.2:* The set of functions $\{f_N\}$ in (67) can be expressed in the form

$$f_N(x) = \sum_{j=0}^{N-1} \exp\big(d(N,j,x)\big) \ , \quad x \in \mathbb{R}, \ N \in \mathbb{N}$$

where

$$
\begin{aligned}
d(N,j,x) \ \triangleq \ & \frac{x^2}{2} + \ln \Gamma\left(\frac{N}{2}\right) - \ln \Gamma\left(\frac{j}{2}+1\right) - \ln \Gamma(N-j) \\
& + (N-1-j)\ln\left(\sqrt{2}\,x\right) - \frac{\ln 2}{2} \\
& + \ln\left[1 + (-1)^j\,\tilde{\gamma}\left(\frac{x^2}{2}, \frac{j+1}{2}\right)\right] \ , \qquad
\begin{array}{l} N \in \mathbb{N}, \ x \in \mathbb{R} \\ j = 0, 1 \dots, N-1 \end{array}
\end{aligned}
\tag{74}
$$

and

$$\Gamma(a) \ \triangleq \ \int_0^\infty t^{a-1} e^{-t} dt \ , \quad \text{Re}(a) > 0 \tag{75}$$

$$\tilde{\gamma}(x,a) \ \triangleq \ \frac{1}{\Gamma(a)} \int_0^x t^{a-1} e^{-t} dt \ , \quad x \in \mathbb{R}, \ \text{Re}(a) > 0 \tag{76}$$

designate the complete and incomplete Gamma functions, respectively.

*Proof:* The proof is given in Appendix C. ∎

*Remark 4.1:* It is noted that the exponents $d(N,j,x)$ in (74) are readily calculated by using standard mathematical functions. The function which calculates the natural logarithm of the Gamma function is implemented in the MATLAB software by `gammaln`, and in the Mathematica software by `LogGamma`. The function $\tilde{\gamma}(a,b)$ is implemented in MATLAB by `gammainc(x,N)` and in Mathematica by `GammaRegularized(N,0,x)`.

In order to perform the entire calculation of the function $f_N$ in the log domain, we employ the function

$$\max{}^*(x_1,\dots,x_m) \triangleq \ln\left(\sum_{i=1}^m e^{x_i}\right) \ , \quad m \in \mathbb{N}, \quad x_1,\dots,x_m \in \mathbb{R} \tag{77}$$

which is commonly used in the implementation of the log-domain BCJR algorithm. The function $\max{}^*$ can be calculated in the log domain using the recursive equation

$$\max{}^*(x_1,\dots,x_{m+1}) = \max{}^*\big(\max{}^*(x_1,\dots,x_m), x_{m+1}\big) \ , \quad m \in \mathbb{N} \setminus \{1\}, \ x_1,\dots,x_{m+1} \in \mathbb{R}$$

with the initial condition

$$\max{}^*(x_1,x_2) = \max(x_1,x_2) + \ln\left(1 + e^{-|x_1-x_2|}\right) \ .$$

Combining Proposition 4.2 and the definition of the function $\max{}^*$ in (77), we get a method of calculating the set of functions $\{f_N\}$ in the log domain.

*Corollary 4.2:* The set of functions $\{f_N\}$ defined in (67) can be rewritten in the form

$$f_N(x) = \exp\left[\max{}^*\big(d(N,0,x), d(N,1,x), \dots, d(N,N-1,x)\big)\right] \tag{78}$$

where $d(N,j,x)$ is introduced in (74).

By combining (66) and (78), one gets the following theorem which provides an efficient algorithm for the calculation of the SP59 bound in the log domain.

*Theorem 4.4 (Log domain calculation of the SP59 bound):* The term $P_{\text{SPB}}(N,\theta,A)$ in the RHS of (70) can be rewritten as

$$
\begin{aligned}
P_{\text{SPB}}(N,\theta,A) \ = \ & \int_\theta^{\frac{\pi}{2}} \exp\left[\ln(N-1) - \frac{NA^2}{2} - \frac{1}{2}\ln(2\pi) + (N-2)\ln\sin\phi \right. \\
& \left. + \max{}^*\big(d(N,0,\sqrt{N}A\cos\phi), \dots, d(N,N-1,\sqrt{N}A\cos\phi)\big)\right] d\phi \\
& + Q(\sqrt{N}A) \ , \qquad N \in \mathbb{N}, \ \theta \in [0, \tfrac{\pi}{2}], \ A \in \mathbb{R}^+
\end{aligned}
$$

where $d(N,j,x)$ is defined in (74).

Using Theorem 4.4, it is easy to calculate the exact value of the SP59 lower bound for very large block lengths.

## V. Numerical Results for Sphere-Packing Bounds

This section presents some numerical results which serve to demonstrate the improved tightness of the ISP bound derived in Section III. We consider performance bounds for M-ary PSK block coded modulation with coherent detection over an AWGN channel, and for the binary erasure channel (BEC) which is MBIOS. As noted in Section III-A, these channels are symmetric and hence the ISP bound holds in these cases. For M-ary PSK modulated signals transmitted over the AWGN channel, the ISP bound is also compared with the SP59 bound revisited in Section IV and some upper bounds on the decoding error probability. We also compare these bounds to some computer simulations of iteratively decoded codes, and examine the tightness of these bounds w.r.t. the performance of modern error-correcting codes using practical decoding algorithms.

### A. Performance Bounds for M-ary PSK Block Coded Modulation over the AWGN Channel

The ISP bound in Section III is particularized here to M-ary PSK block coded modulation schemes whose transmission takes place over an AWGN channel, and where the received signals are coherently detected. For simplicity of notation, we treat the channel inputs and outputs as two dimensional real vectors, and not as complex numbers. Let $M = 2^p$ (where $p \in \mathbb{N}$) be the modulation parameter, denote the input to the channel by $\mathbf{X} = (x_1, x_2)$ where the possible input values are given by

$$\mathbf{X}_k = (\cos \theta_k, \sin \theta_k), \quad \theta_k \triangleq \frac{(2k+1)\pi}{M}, \quad k = 0, 1, \ldots, M-1. \tag{79}$$

We denote the channel output by $\mathbf{Y} = (y_1, y_2)$ where $\mathbf{Y} = \mathbf{X} + \mathbf{N}$, and $\mathbf{N} = (n_1, n_2)$ is a Gaussian random vector with i.i.d. components each with zero-mean and variance $\sigma^2$. The conditional *pdf* of the channel output, given the transmitted symbol $\mathbf{X}_k$, is given by

$$p_{\mathbf{Y}|\mathbf{X}}(\mathbf{Y}|\mathbf{X}_k) = \frac{1}{2\pi\sigma^2} e^{-\frac{\|\mathbf{Y} - \mathbf{X}_k\|^2}{2\sigma^2}}, \quad \mathbf{Y} \in \mathbb{R}^2 \tag{80}$$

where $\|\cdot\|$ designates the $L_2$ norm. The closed form expressions for the function $\mu_0$ and its first two derivatives w.r.t. $s$ (while holding $f_s$ fixed) are derived in Appendix B.1 and are used for the calculation of both the VF and ISP bounds. The SP59 bound [24] provides a lower bound on the decoding error probability for the considered case, since the modulated signals have equal energy and are transmitted over the AWGN channel. In the following, we exemplify the use of these lower bounds. They are also compared to the random-coding upper bound of Gallager [11], and the tangential-sphere upper bound (TSB) of Poltyrev [20] when applied to random block codes. This serves for the study of the tightness of the ISP bound, as compared to other upper and lower bounds. The numerical results shown in this section indicate that the recent variants of the SP67 bound provide an interesting alternative to the SP59 bound which is commonly used in the literature as a measure for the sub-optimality of codes transmitted over the AWGN channel (see, e.g., [9], [14], [17], [23], [29], [35], [37]). Moreover, the advantage of the ISP bound over the VF bound in [35] is exemplified in this section.

Figure 2 compares the SP59 bound [24], the VF bound [35], and the ISP bound derived in Section III. The comparison refers to block codes of length 500 bits and rate $0.8 \frac{\text{bits}}{\text{channel use}}$ which are BPSK modulated and transmitted over an AWGN channel. The plot also depicts the random-coding bound of Gallager [11], the TSB ([13], [20]), and the capacity limit bound (CLB).[1] It is observed from this figure that even for relatively short block lengths, the ISP bound outperforms the SP59 bound for block error probabilities below $10^{-1}$ (this issue will be discussed later in this section). For a block error probability of $10^{-5}$, the ISP bound provides gains of about 0.26 and 0.33 dB over the SP59 and VF bounds, respectively. For these code parameters, the TSB provides a tighter upper bound on the block error probability of random codes than the random-coding bound; e.g., the gain of the TSB over the Gallager bound is about 0.2 dB for a block error probability of $10^{-5}$. Note that the Gallager bound is tighter than the TSB for fully random block codes of large enough block lengths, as the latter bound does not reproduce the random-coding error exponent for the AWGN channel [20]. However, Figure 2 exemplifies the advantage of the TSB over the Gallager bound, when applied to random block codes of relatively short block lengths; this advantage is especially pronounced for low code rates where the gap between the error exponents of these two bounds is marginal (see [23, p. 67] and [32]), but it is also reflected from Figure 2 for BPSK modulation with a code rate of $0.8 \frac{\text{bits}}{\text{channel use}}$. The gap between the TSB and the ISP bound, as upper and lower bounds respectively, is less than 1.2 dB for all block error probabilities lower than $10^{-1}$. Also, the ISP bound is more informative than the CLB for block error probabilities below $8 \cdot 10^{-3}$ while the SP59 and VF bounds require block error probabilities below $1.5 \cdot 10^{-3}$ and $5 \cdot 10^{-4}$, respectively, to outperform the capacity limit.

---

[1]Although the CLB refers to the asymptotic case where the block length tends to infinity, it is plotted in [35] and here as a reference, in order to examine whether the improvement in the tightness of the ISP is for rates above or below capacity.

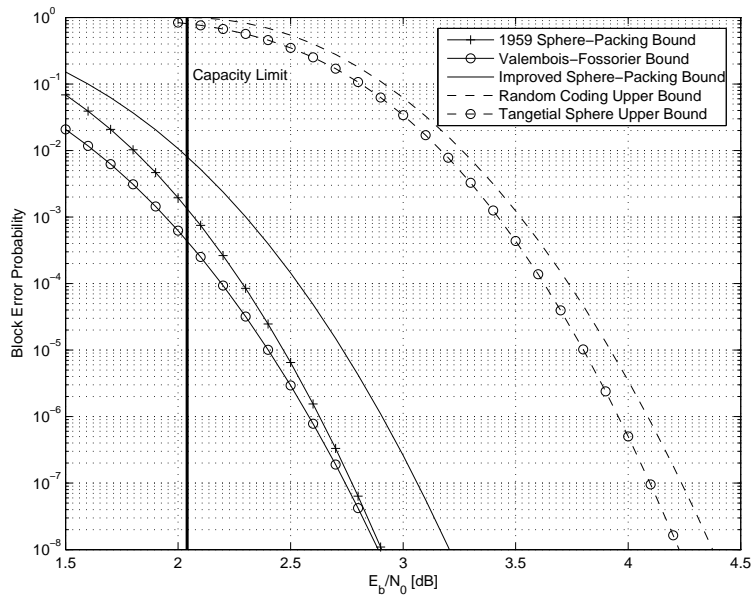

Fig. 2. A comparison between upper and lower bounds on the ML decoding error probability for block codes of length $N = 500$ bits and code rate of $0.8 \frac{\text{bits}}{\text{channel use}}$. This figure refers to BPSK modulated signals whose transmission takes place over an AWGN channel. The compared bounds are the 1959 sphere-packing (SP59) bound of Shannon [24], the Valembois-Fossorier (VF) bound [35], the improved sphere-packing (ISP) bound derived in Section III, the random-coding upper bound of Gallager [11], and the TSB [13], [20] when applied to fully random block codes with the above block length and rate.

Figure 3 presents a comparison of the SP59, VF and ISP bounds referring to short block codes which are QPSK modulated and transmitted over the AWGN channel. The plots also depict the random-coding upper bound, the TSB and CLB; in these plots, the ISP bound outperforms the SP59 bound for all block error probabilities below $4 \cdot 10^{-1}$ (this result is consistent with the upper plot of Figure 7). In the upper plot of Figure 3, which corresponds to a block length of 1024 bits (i.e., 512 QPSK symbols) and a rate of $1.5 \frac{\text{bits}}{\text{channel use}}$, it is shown that the ISP bound provides gains of about 0.25 and 0.37 dB over the SP59 and VF bounds, respectively, for a block error probability of $10^{-5}$. The gap between the ISP lower bound and the random-coding upper bound is 0.78 dB for all block error probabilities lower than $10^{-1}$. In the lower plot of Figure 3 which corresponds to a block length of 300 bits and a rate of $1.8 \frac{\text{bits}}{\text{channel use}}$, the ISP bound significantly improves the SP59 and VF bounds; for a block error probability of $10^{-5}$, the improvement in the tightness of the ISP over the SP59 and VF bounds is 0.8 and 1.13 dB, respectively. Additionally, the ISP bound is more informative than the CLB for block error probabilities below $3 \cdot 10^{-3}$, where the SP59 and VF bound outperform the CLB only for block error probabilities below $3 \cdot 10^{-6}$ and $5 \cdot 10^{-8}$, respectively. For fully random block codes of length $N = 300$ and rate $1.8 \frac{\text{bits}}{\text{channel use}}$ which are QPSK modulated with Gray's mapping and transmitted over the AWGN channel, the TSB is tighter than the random-coding bound (see the lower plot in Figure 3 and the explanation referring to Figure 2). The gap between the ISP bound and the TSB in this plot is about 1.5 dB for a block error probability of $10^{-5}$ (as compared to gaps of 2.3 dB (2.63 dB) between the TSB and the SP59 (VF) bound).

Figure 4 presents a comparison of the bounds for codes of block length 5580 bits and 4092 information bits, where both QPSK (upper plot) and 8-PSK (lower plot) constellations are considered. The modulated signals correspond to 2790 and 1680 symbols, respectively, so the code rates for these constellations are 1.467 and 2.2 bits per channel use, respectively. For both constellations, the two considered SP67-based bounds (i.e., the VF and ISP bounds) outperform the SP59 for all block error probabilities below $2 \cdot 10^{-1}$; the ISP bound provides gains of 0.1 and 0.22 dB over the VF bound for the QPSK and 8-PSK constellations, respectively. For both modulations, the gap between the ISP lower bound and the random-coding upper bound of Gallager does not exceed $0.4$ dB. In [6], Divsalar and Dolinar design codes with the considered parameters by using concatenated Hamming and accumulate codes. They also present computer simulations of the performance of these codes under iterative decoding, when the transmission takes place over the AWGN channel and several common modulation schemes are applied. For a block error probability of $10^{-4}$, the gap between the simulated performance of these codes under iterative decoding, and the ISP lower bound, which gives an ultimate lower bound on the block error probability of optimally designed codes under ML decoding, is approximately $1.4$ dB for QPSK and $1.6$ dB for 8-PSK signaling. This provides an indication on the performance of codes defined on graphs and their iterative decoding algorithms, especially in light of the feasible complexity of the decoding algorithm which is linear in the block length. To conclude, it is reflected from the results plotted in Figure 4 that a gap of about 1.5 dB between the ISP lower bound and the performance of the iteratively decoded codes in [6] is mainly due to the imperfectness of these codes and their sub-optimal iterative decoding algorithm; this conclusion follows in light of the fact that for random codes of the same block length and rate, the gap between the ISP bound and the random coding bound is reduced to less than 0.4 dB.

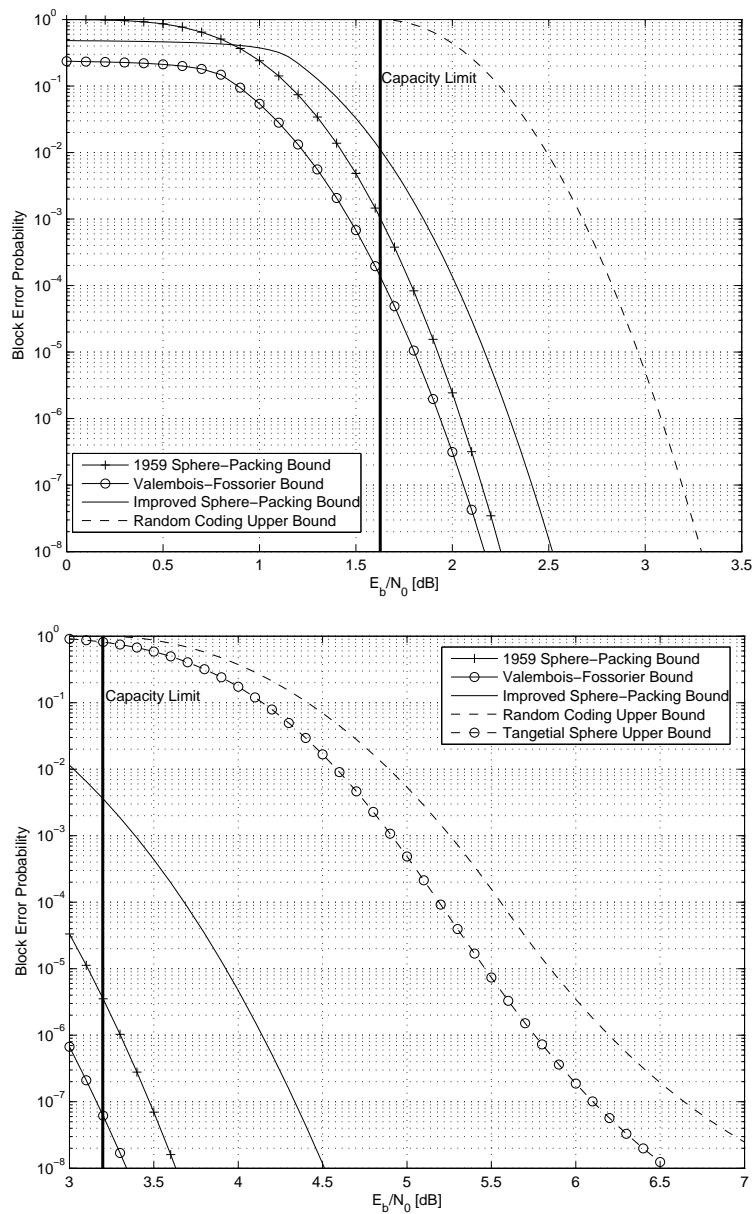

Fig. 3. A comparison between upper and lower bounds on the ML decoding error probability, referring to short block codes which are QPSK modulated and transmitted over the AWGN channel. The compared lower bounds are the 1959 sphere-packing (SP59) bound of Shannon [24], the Valembois-Fossorier (VF) bound [35], and the improved sphere-packing (ISP) bound; the compared upper bounds are the random-coding upper bound of Gallager [11] and the tangential-sphere bound (TSB) of Poltyrev [20]. The upper plot refers to block codes of length $N = 1024$ which are encoded by 768 information bits (so the rate is $1.5 \frac{\text{bits}}{\text{channel use}}$), and the lower plot refers to block codes of length $N = 300$ which are encoded by 270 bits whose rate is therefore $1.8 \frac{\text{bits}}{\text{channel use}}$.

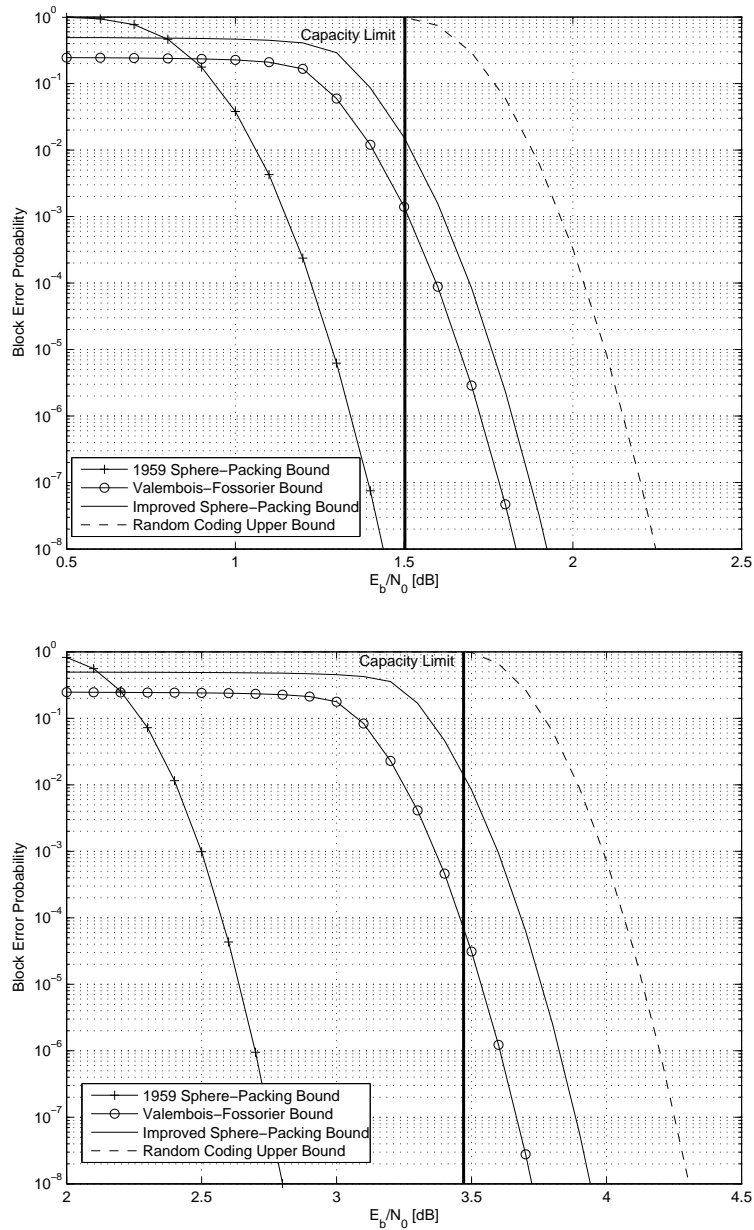

Fig. 4. A comparison of upper and lower bounds on the ML decoding error probability for block codes of length $N = 5580$ bits and information block length of 4092 bits. This figure refers to QPSK (upper plot) and 8-PSK (lower plot) modulated signals whose transmission takes place over an AWGN channel; the rates in this case are 1.467 and 2.200 $\frac{\text{bits}}{\text{channel use}}$, respectively. The compared bounds are the 1959 sphere-packing (SP59) bound of Shannon [24], the Valembois-Fossorier (VF) bound [35], the improved sphere-packing (ISP) bound, and the random-coding upper bound of Gallager [11].

While it was shown in Section III that the ISP bound is uniformly tighter than the VF bound (which in turn is uniformly tighter than the SP67 bound [25]), no such relations are shown between the SP59 bound and the recent improvements on the SP67 bound (i.e., the VF and ISP bounds). Figure 5 presents regions of code rates and block lengths for which the ISP bound outperforms the SP59 bound and the CLB; it refers to BPSK modulated signals transmitted over the AWGN channel and considers block error probabilities of $10^{-4}$, $10^{-5}$ and $10^{-6}$. It is reflected from this figure that for any rate $0 < R < 1$, there exists a block length $N = N(R)$ such that the ISP bound outperforms the SP59 bound for block lengths larger than $N(R)$; the same property also holds for the VF bound, but the value of $N(R)$ depends on the considered SP67-based bound, and it becomes significantly larger in the comparison of the VF and SP59 bounds. It is also observed that the value $N(R)$ is monotonically decreasing with $R$, and it approaches infinity as we let $R$ tend to zero. An intuitive explanation for this behavior can be given by considering the capacity limits of the binary-input and the energy-constrained AWGN channels. For any value $0 \leq C < 1$, denote by $\frac{E_{b,1}(C)}{N_0}$ and $\frac{E_{b,2}(C)}{N_0}$ the values of $\frac{E_b}{N_0}$ required to achieve a channel capacity of $C$ bits per channel use for the binary-input and the energy-constraint AWGN channels, respectively (note that in the latter case, the input distribution which achieves capacity is also Gaussian). For any $0 \leq C < 1$, clearly $\frac{E_{b,1}(C)}{N_0} \geq \frac{E_{b,2}(C)}{N_0}$; however, the difference between these values is monotonically increasing with the capacity $C$, and, on the other hand, this difference approaches zero as we let $C$ tend to zero. Since the SP59 bound only constrains the signals to be of equal energy, it gives a measure of performance for the energy-constrained AWGN channel, where the SP67-based bounds consider the actual modulation and therefore refer to the binary-input AWGN channel. As the code rates become higher, the difference in the ultimate performance between the two channels is larger, and therefore the SP67-based bounding techniques outperform the SP59 bound for smaller block lengths. For low code rates, the difference between the channels is reduced, and the SP59 outperforms the SP67-based bounding techniques even for larger block lengths due to the superior bounding technique which is specifically tailored for the AWGN channel. Figure 6 presents the regions of code rates and block lengths for which the VF bound (left plot) and the ISP bound (right plot) outperform the CLB and the SP59 bound when the signals are BPSK modulated and transmitted over the AWGN channel; block error probabilities of $10^{-4}$, $10^{-5}$ and $10^{-6}$ are examined. This figure is focused on high code rates, where the performance of the SP67-based bounds and their advantage over the SP59 bound is most appealing. From Figure 6, we have that for a code rate of 0.75 bits per channel use and a block error probability of $10^{-6}$, the VF bound becomes tighter than the SP59 for block lengths exceeding 850 bits while the ISP bound reduces this value to 450 bits; moreover, when increasing the rate to 0.8 bits per channel use, the respective minimal block lengths reduce to 550 and 280 bits for the VF and ISP bounds, respectively. Fig 7 shows the regions of code rates and block lengths where the ISP outperforms the CLB and SP59 bounds for QPSK (left plot) and 8-PSK (right plot) modulations. Comparing the lower plot of Figure 6 which refers to BPSK modulation with the upper plot of Figure 7 which refers to QPSK modulation, one can see that the two graphs are identical (when accounting for the doubling of the rate which is due to the use of both real and imaginary dimensions in the QPSK modulation). This is due to the fact that QPSK modulation poses no additional constraints on the channel and in fact, the real and imaginary planes can be serialized and decoded as in BPSK modulation. However, this property does not hold when replacing the ISP bound by the VF bound; this is due to the fact that the VF bound considers a fixed composition subcode of the original code and the increased size of the alphabet causes a greater loss in the rate for QPSK modulation. When comparing the two plots of Figure 7, it is evident that the minimal value of the block length for which the ISP bound becomes better than the SP59 bound decreases as the size of the input alphabet is increased (when the rate is measured in units of information bits per code bit). An intuitive justification for this phenomenon is attributed to the fact that referring to the constellation points of the M-ary PSK modulation, the mutual information between the code symbols in each dimension of the QPSK modulation is zero, while as the spectral efficiency of the PSK modulation is increased, the mutual information between the real and imaginary parts of each signal point is increased; thus, as the spectral efficiency is increased, this poses a stronger constraint on the possible positioning of the equal-energy signal points on the $N$-dimensional sphere. This intuition suggests an explanation for the reason why as the spectral efficiency is increased, the advantage of the ISP bound over the SP59 bound (where the latter does not take into account the modulation scheme) holds even for smaller block lengths. This effect is expected to be more subtle for the VF bound since a larger size of the input alphabet decreases the rate for which the error exponent is evaluated (see (22)).

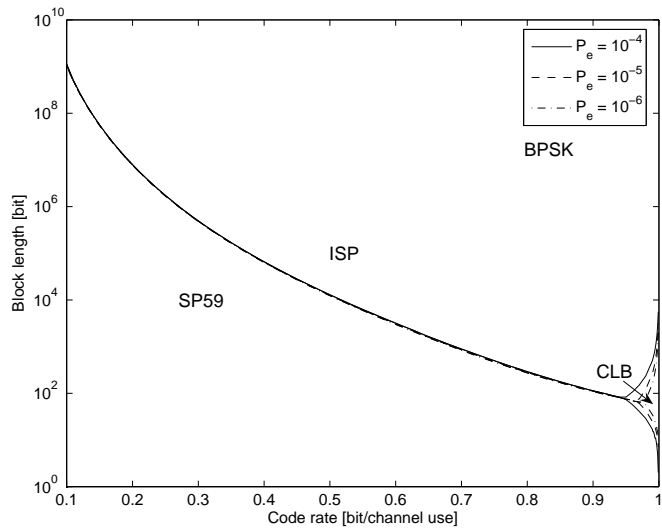

Fig. 5. Regions in the two-dimensional space of code rate and block length, where a bound is better than the two others for three different targets of block error probability ($P_e$). The figure compares the tightness of the 1959 sphere-packing (SP59) bound of Shannon [24], the improved sphere-packing (ISP) bound, and the capacity-limit bound (CLB). The plot refers to BPSK modulated signals whose transmission takes place over the AWGN channel, and the considered code rates lie in the range between 0.1 and $1 \frac{\text{bits}}{\text{channel use}}$.

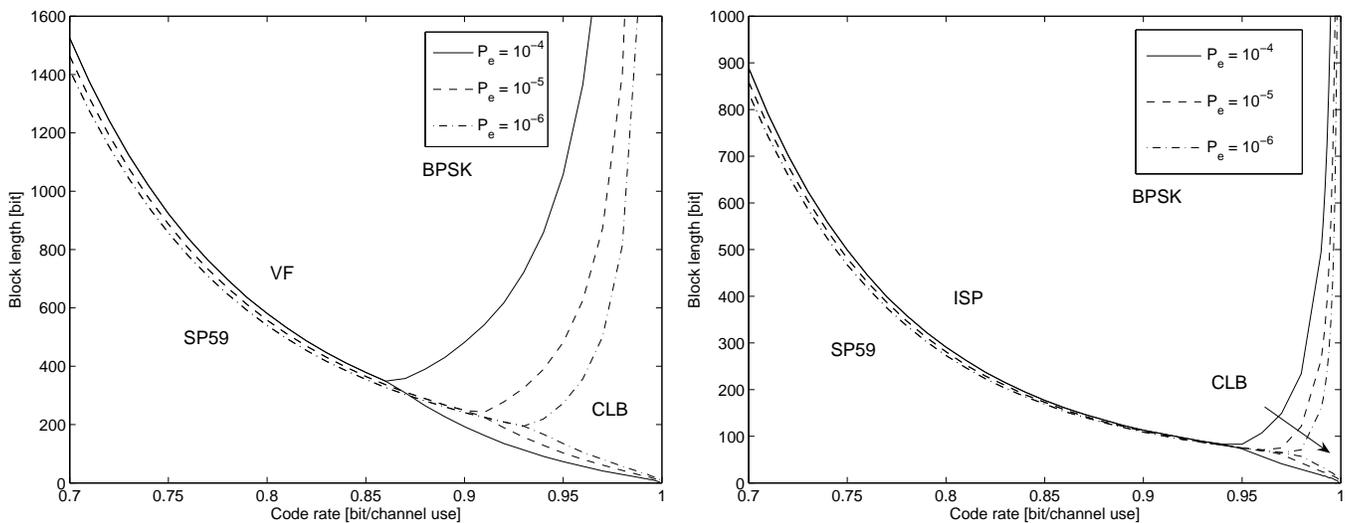

Fig. 6. Regions in the two-dimensional space of code rate and block length, where a bound is better than the two others for three different targets of block error probability ($P_e$). The figure compares the tightness of the 1959 sphere-packing (SP59) bound of Shannon [24], the capacity-limit bound (CLB), and the Valembois-Fossorier (VF) bound [35] (left plot) or the improved sphere-packing (ISP) bound in Section III (right plot). The plots refer to BPSK modulated signals whose transmission takes place over the AWGN channel, and the considered code rates lie in the range between 0.70 and $1 \frac{\text{bits}}{\text{channel use}}$.

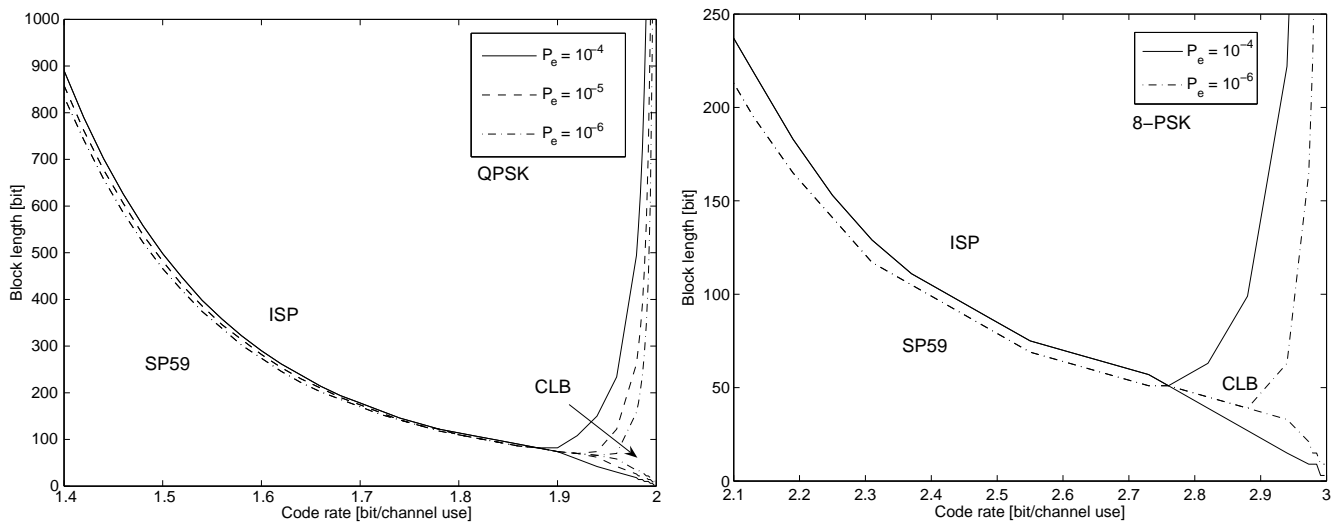

Fig. 7. Regions in the two-dimensional space of code rate and block length, where a bound is better than the two others for different targets of block error probability ($P_e$). The figure compares the tightness of the 1959 sphere-packing (SP59) bound of Shannon [24], the improved sphere-packing (ISP) bound, and the capacity-limit bound (CLB). The plots refer to QPSK (left plot) and 8-PSK (right plot) modulated signals whose transmission takes place over the AWGN channel; the considered code rates lie in the range between 1.4 and $2 \frac{\text{bits}}{\text{channel use}}$ for the QPSK modulated signals and between 2.1 and $3 \frac{\text{bits}}{\text{channel use}}$ for the 8-PSK modulated signals.

*B. Performance Bounds for the Binary Erasure Channel*

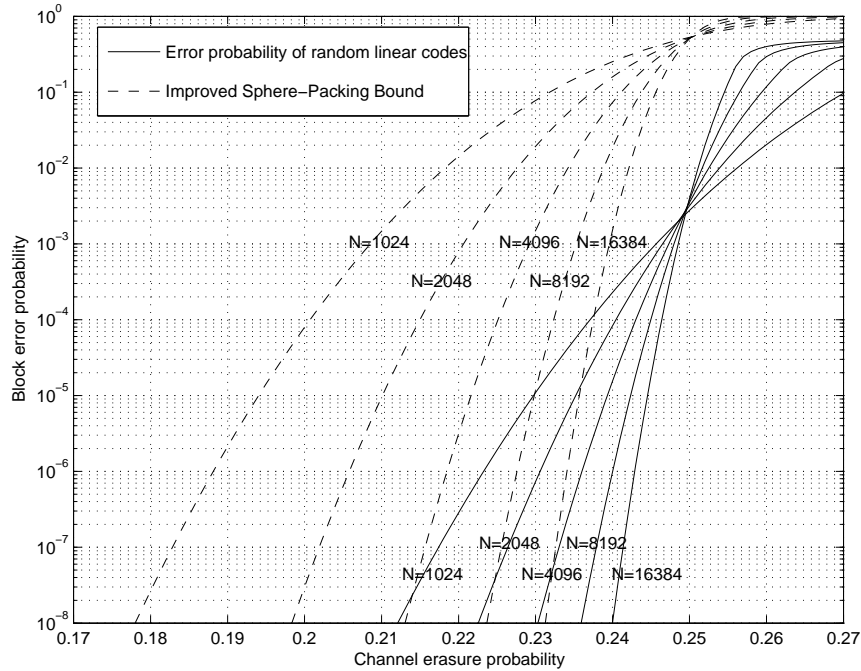

Fig. 8. A comparison of the improved sphere-packing (ISP) lower bound from Section III and the exact decoding error probability of random binary linear block codes under ML decoding where the transmission takes place over the BEC (see [5, Eq. (3.2)]). The code rate examined is $0.75 \frac{\text{bits}}{\text{channel use}}$ and the block lengths are $N = 1024, 2048, 4096, 8192$ and $16384$ bits.

In recent years, the BEC has been the focus of much attention in the field of iterative decoding techniques. The simplicity of this channel and the absolute reliability of the known values at the output lend themselves to a one-dimensional analysis of turbo-like codes and the performance of their iterative decoding algorithms in the case where the codes are transmitted over the BEC (see, e.g., [27]). For the asymptotic case where we let the block length tend to infinity, several families which achieve the capacity of the BEC under iterative decoding have been constructed; these include low-density parity-check (LDPC) [16], irregular repeat-accumulate (RA) [18] and accumulate-repeat-accumulate (ARA) codes [19]; several ensembles of IRA and ARA codes were demonstrated to achieve the capacity of the BEC with bounded complexity per information bit, in contrast to LDPC codes without puncturing whose decoding complexity necessarily becomes unbounded as the gap to capacity vanishes (see [22], [18], [19]). These discoveries motivate a study of the performance of iteratively decoded codes defined on graphs for moderate block lengths (see, e.g., [33]). In Figure 8, we compare the ISP lower bound and the exact block error probability of random linear block codes transmitted over the BEC as given in [5, Eq. (3.2)]. The figure refers to codes of rate 0.75 bits per channel use and various block lengths. It can be observed that for a block length of 1024 bits, the difference in the channel erasure probability for which the random coding bound and the ISP bound achieve a block error probability of $10^{-5}$ is 0.035 while for a block length of 16384 bits, this gap is decreased to 0.009. This yields that the ISP bound is reasonably tight, and also suggests that this bound can be used in order to assess the imperfectness of turbo-like codes even for moderate block lengths.

*C. Minimal Block Length as a Function of Performance*

In a wide range of applications, the system designer needs to design a communication system which fulfills several requirements on the available bandwidth, acceptable delay for transmitting and processing the data while maintaining a certain fidelity criterion in reconstructing the data (e.g., the block error probability needs to be below a certain threshold). In this setting, one wishes to design a block code which satisfies the delay constraint (i.e., the block length is limited) while adhering to the required performance over the given channel. By fixing the communication channel model, code rate (which is related to the bandwidth expansion caused by the error-correcting code) and the block error probability, sphere-packing bounds are transformed into lower bounds on the minimal block length required to achieve the desired block error probability at a certain gap to capacity using an arbitrary block code and decoding algorithm. Similarly, by fixing these parameters, the random coding bound of Gallager [11] is transformed into an upper bound on the block length required for ML decoded random codes to achieve a desired block error probability on a given communication channel.

In this section, we consider some practically decodable codes taken from some recent papers ([1], [7], [8], [28], [30], [34]). We examine the gap between channel capacity and the $\frac{E_b}{N_0}$ for which they achieve a required block error probability as a function of the block length of these codes. The performance of these specific codes together with their practical decoding algorithms is compared with the sphere-packing and random coding bounds; these bounds serve here as lower and upper bounds, respectively, on the block length required to achieve a given block error probability and code rate on a given channel using an optimal block code and decoding algorithm. This comparison shows how far, in terms of delay, some modern error-correcting codes with their sub-optimal and practical decoding algorithms are from the fundamental limitations imposed by information theory.

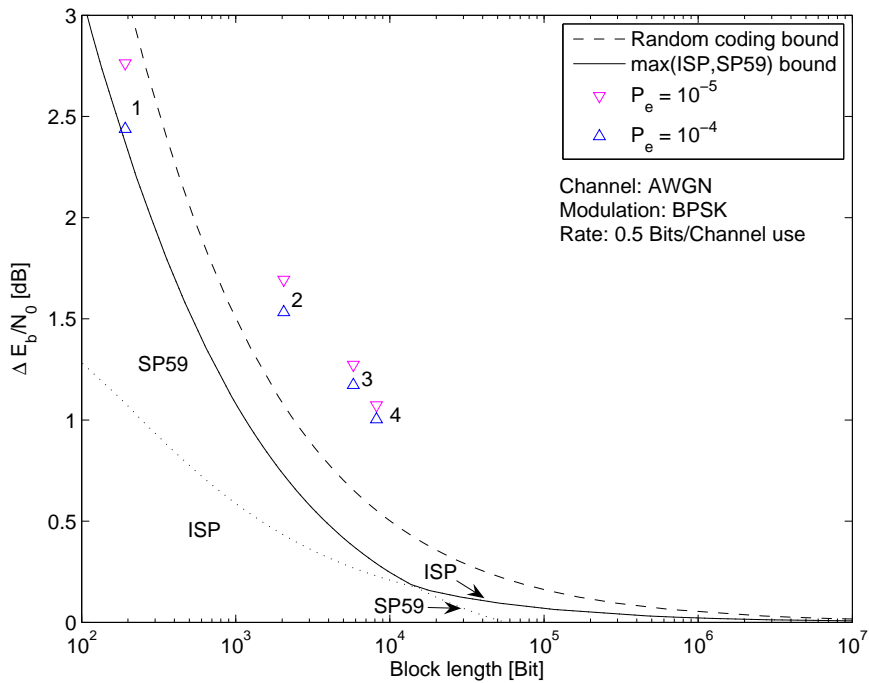

Fig. 9. Bounds on the block length required to achieve a block error probability of $10^{-5}$ compared with the performance of some practically decodable codes. The considered communication channel model is the BPSK modulated AWGN channel and the rate of the codes is 0.5 bits per channel use. The depicted bounds are the 1959 sphere-packing (SP59) bound of Shannon [24], the improved sphere-packing (ISP) bound introduced in Section III, and the random coding upper bound of Gallager [11]. The codes are taken from [34] (code 1), [8] (codes 2 and 4) and [7] (code 3).

Figure 9 considers some block codes of rate $\frac{1}{2}$ bits per channel use which are BPSK modulated and transmitted over the AWGN channel. The plot depicts the gap to capacity in dB for which these codes achieve block error probabilities of $10^{-4}$ and $10^{-5}$ under their practical decoding algorithms as a function of their block length. As a reference, this figure also plots lower bounds on the block length which stem from the SP59 and ISP bounds and the upper bound on the block length of random codes which stems from the random-coding bound of Gallager.

The three bounds refer to a block error probability of $10^{-5}$. The code labeled 1 is a block code of length 192 bits which is decoded using a near-ML decoder by applying 'box and match' decoding techniques [34]. It is observed that this code outperforms random coding upper bound for ML decoded random codes with the same block length and code rate. It is also observed that this code achieves a block error probability of $10^{-5}$ at a gap to capacity of 2.76 dB while the SP59 bound gives that the block length required to achieve this performance is lower bounded by 133 bits (so the bound is very informative). The codes labeled 2, 3 and 4 are prototype-based LDPC codes of lengths 2048, 5176 and 8192 bits, respectively (codes 2 and 4 are taken from [8] and code 3 is taken from [7]). These codes achieve under iterative decoding a block error probability of $10^{-5}$ at gaps to capacity of 1.70, 1.27 and 1.07 dB, respectively. In terms of block length, the gap between the performance of these codes under iterative decoding and the SP59 lower bound on the block length required to achieve a block error probability of $10^{-5}$ at these channel conditions is less than one order of magnitude. It is also noted that throughout the range of block lengths depicted in Figure 9, the gap between the lower bound on the block length of optimal codes which stems from the better of the two sphere-packing bounds and the upper bound on the block length of random codes is less than one order of magnitude. This exemplifies the tightness of the sphere-packing bounds when used as lower bounds on the block lengths of optimal codes.

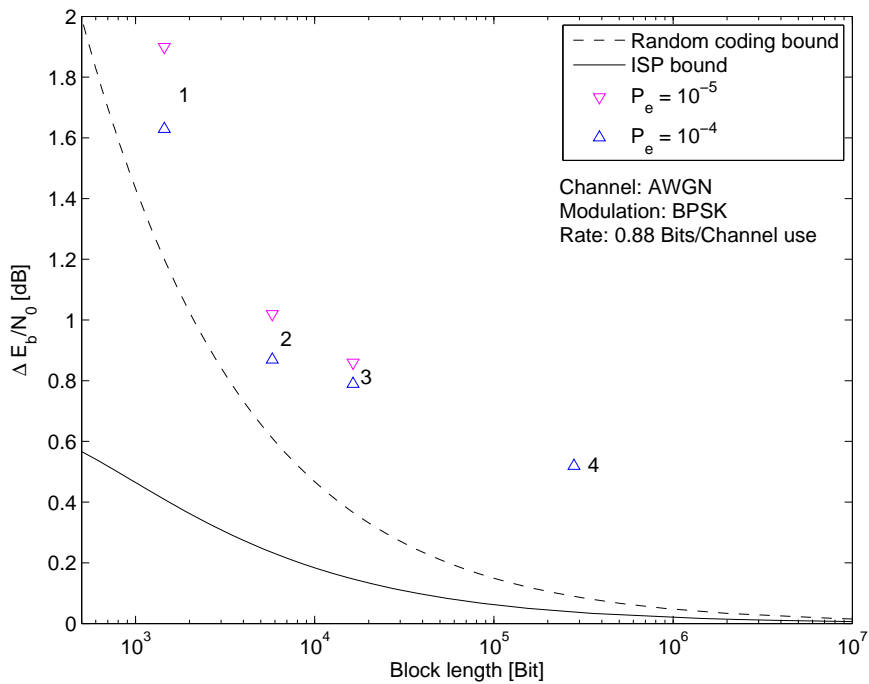

Fig. 10. Bounds on the block length required to achieve a block error probability of $10^{-5}$ compared with the performance of some practically decodable codes. The considered communication channel model is the BPSK modulated AWGN channel and the rate of the codes is 0.5 bits per channel use. The depicted bounds are the improved sphere-packing (ISP) bound introduced in Section III, and the random coding upper bound of Gallager [11]. Codes 1,2,3 and 4 are taken from [1], [7], [28] and [30], respectively.

Figure 10 considers some LDPC codes of rate 0.88 bits per channel use which are BPSK modulated and transmitted over the AWGN channel. The gap to capacity in dB for which these codes achieve block error probabilities of $10^{-4}$ and $10^{-5}$ under iterative decoding is plotted as a function of block length. As in Figure 9, the figure uses lower and upper bounds on the block length which stem from the ISP and random-coding bounds, respectively. For the rate and block lengths depicted, the SP59 bound is universally looser than the ISP bound and hence it is not shown. The bounds refer to a block error probability of $10^{-5}$. For the examined block error probabilities, the depicted codes require a gap to capacity of between 0.63 and 1.9 dB. For this range of $\frac{E_b}{N_0}$ the lower bound on the block lengths which is derived from the ISP bound is looser than the one given by the SP59 bound. However, both bounds are not very informative in this range. For cases where the gap to capacity is below 0.5 dB, the difference between the lower bound on the block length of optimal codes which stems from the ISP bound and the upper bound on the block length of random codes is less than one order of magnitude. Code number 1 is an LDPC of length 1448 bits whose construction of is based on balanced incomplete block designs [1]. This code achieves a block error probability of $10^{-5}$ at a gap to capacity of 1.9 dB while the random-coding bound shows that the block length required to achieve this performance using random codes is upper bounded by 600 bits. The code labeled 2 is a prototype-based LDPC code of length 5176 bits which is taken from [7]. Code number 3 is a quasi-cyclic LDPC code of length 16362 bits taken from [28]. These code achieve under iterative decoding a block error probability of $10^{-5}$ at gaps to capacity of 1.02 and 0.86 dB, respectively. In terms of block length, the gap between the performance of these codes under iterative decoding and the upper bound on the block length of random codes which achieve a block error probability of $10^{-5}$ under the same channel conditions is less than one order of magnitude. The code labeled 4 is a finite-geometry LDPC code of length 279552 bits which is taken from [30]. For this code we only have the gap to capacity required to achieve a block error probability of $10^{-4}$, however, it is clear that the difference in block length from the random coding upper bound becomes quite large as the gap to capacity is reduced.

By fixing the block length and considering the gap in $\Delta E_{\rm b}/N_0$ between the performance of the specific codes and the sphere-packing bounds in Figures 9 and 10, it is observed that the codes considered in these plots achieve exhibit gaps of $0.2 - 0.6$ dB w.r.t. the information-theoretic limitation provided by the sphere-packing bounds (with the exception of code 1 in Figure 10 which exhibits a gap of about 1.25 dB). In this respect we also mention some high rate turbo-product codes with moderate block lengths (see [4]) exhibit a gap of $0.75 - 0.95$ dB w.r.t. the information-theoretic limitation provided by the ISP bound. Based on numerical results in [31] for the ensemble of uniformly interleaved (1144, 1000) turbo-block codes whose components are random systematic, binary and linear block codes, the gap in $\frac{E_{\rm b}}{N_0}$ between the ISP lower bound and an upper bound under ML decoding is 0.9 dB for a block error probability of $10^{-7}$. These results exemplify the strength of the sphere-packing bounds for assessing the theoretical limitations of block codes and the power of iteratively decoded codes (see also [9], [14], [15], [23], [35]).

## VI. Summary

This paper presents an improved sphere-packing (ISP) bound for finite-length block codes whose transmission takes place over symmetric memoryless channels. The improvement in the tightness of the bound is especially pronounced for codes of short to moderate block lengths, and some of its applications are exemplified in this paper. The derivation of the ISP bound was stimulated by the remarkable performance and feasible complexity of turbo-like codes with short to moderate block lengths. We were motivated by recent improvements on the sphere-packing bound of [25] for finite block lengths, as suggested by Valembois and Fossorier [35].

We first review the classical sphere-packing bounds, i.e., the 1959 sphere-packing bound (SP59) derived by Shannon for equal-energy signals transmitted over the Gaussian channel [24], and the 1967 sphere-packing (SP67) bound derived by Shannon, Gallager and Berlekamp for discrete memoryless channels [25]. The ISP bound, introduced in Section III, is uniformly tighter than the classical SP67 bound [25] and the bound in [35].

We apply the ISP bound to M-ary PSK block coded modulation schemes whose transmission takes place over an AWGN channel and the received signals are coherently detected. The tightness of the ISP bound is exemplified by comparing it with upper and lower bounds on the ML decoding error probability and also with reported computer simulations of turbo-like codes under iterative decoding. The paper also presents a new algorithm which performs the entire calculation of the SP59 bound in the logarithmic domain, thus facilitating the exact calculation of the SP59 bound for all block lengths without the need for asymptotic approximations. It is shown that the ISP bound suggests an interesting alternative to the SP59 bound, where the latter is specialized for the AWGN channel.

In a wide range of applications, one wishes to design a block code which satisfies a known delay constraint (i.e., the block length is limited) while adhering to a required performance over a given channel model. By fixing the communication channel model, code rate and the block error probability, sphere-packing bounds are transformed into lower bounds on the minimal block length required to achieve the desired block error probability at a certain gap to capacity using an arbitrary block code and decoding algorithm. Comparing the performance of specific codes and decoding algorithms to the information-theoretic limitations provided by the sphere-packing bounds, enables one to deduce how far in terms of delay is a practical system from the fundamental limitations of information-theory. Further details on the comparison between practically decodable codes and the sphere-packing bounds are found in Section V-C.

The ISP bound is especially attractive for block codes of short to moderate block lengths; this is especially pronounced for high rate codes (see Figs. 3–7). Its improvement over the SP67 bound and the bound in [35, Theorem 7] is also more significant as the input alphabet of the considered modulation is increased.

*Acknowledgment*

The authors are grateful to Prof. Amos Lapidoth for raising the question which stimulated the discussion in Section V-C.

## Appendices

### Appendix A: Proof of Lemma 3.1

We consider a symmetric DMC with input alphabet $\mathcal{K} = \{0, \ldots, K-1\}$, output alphabet $\mathcal{J} = \{0, \ldots J-1\}$ (where $J, K \in \mathbb{N}$) and transition probabilities $P$. Let $\{g_k\}_{k=0}^{K}$ be the set of unitary functions which satisfy the conditions (24) and (25) in Definition 3.2. To prove Lemma 3.1, we start with a discussion on the distribution $\mathbf{q}_s$ which satisfies (48).

*a) On the input distribution $\mathbf{q}_s$ for symmetric DMCs:*

*Lemma A.1:* For symmetric DMCs and an arbitrary value of $s \in (0,1)$, the uniform distribution $q_{k,s} = \frac{1}{K}$ for $k \in \mathcal{K}$ satisfies (48) in equality.

*Proof:* To prove the lemma, it is required to show that

$$\sum_{j=0}^{J-1} \left\{ P(j|k)^{1-s} \left( \sum_{k'=0}^{K-1} \frac{1}{K} P(j|k')^{1-s} \right)^{\frac{s}{1-s}} \right\} = \sum_{j=0}^{J-1} \left( \sum_{k'=0}^{K-1} \frac{1}{K} P(j|k')^{1-s} \right)^{\frac{1}{1-s}}, \quad \forall k \in \mathcal{K}. \tag{A.1}$$

Let us consider some $k \in \mathcal{K}$. Examining the left-hand side (LHS) of (A.1) gives

$$\begin{aligned}
&\sum_{j=0}^{J-1} \left\{ P(j|k)^{1-s} \left( \sum_{k'=0}^{K-1} \frac{1}{K} P(j|k')^{1-s} \right)^{\frac{s}{1-s}} \right\} \\
&= K \sum_{j=0}^{J-1} \left\{ \frac{1}{K} P(j|k)^{1-s} \left( \sum_{k'=0}^{K-1} \frac{1}{K} P(j|k')^{1-s} \right)^{\frac{s}{1-s}} \right\} \\
&\stackrel{(a)}{=} \sum_{\widetilde{k}=0}^{K-1} \sum_{j=0}^{J-1} \left\{ \frac{1}{K} P(j|k)^{1-s} \left( \sum_{k'=0}^{K-1} \frac{1}{K} P(j|k')^{1-s} \right)^{\frac{s}{1-s}} \right\} \\
&\stackrel{(b)}{=} \sum_{\widetilde{k}=0}^{K-1} \sum_{j=0}^{J-1} \left\{ \frac{1}{K} P(g_{\widetilde{k}}(j)|k)^{1-s} \left( \sum_{k'=0}^{K-1} \frac{1}{K} P(g_{\widetilde{k}}(j)|k')^{1-s} \right)^{\frac{s}{1-s}} \right\}
\end{aligned} \tag{A.2}$$

where $(a)$ holds by summing over a dummy variable $\widetilde{k} \in \{0, 1, \ldots, K-1\}$ instead of the multiplication by $K$ in the previous line, and $(b)$ holds since $g_{\widetilde{k}}$ is unitary for all $\widetilde{k} \in \{0, 1, \ldots, K-1\}$ (see (23) where the integral is replaced here by a sum). For all $j \in \mathcal{J}$ and $\widetilde{k} \in \{0, 1, \ldots, K-1\}$, the symmetry properties in (24) - (26) give

$$\begin{aligned}
P\left(g_{\widetilde{k}}(j)|\, k\right) &\stackrel{(a)}{=} P\left((g_k^{-1} \circ g_{\widetilde{k}})(j)|\, 0\right) \\
&\stackrel{(b)}{=} P\left(g_{(\widetilde{k}-k)\mathrm{mod}K}(j)|\, 0\right) \\
&\stackrel{(c)}{=} P\left(j|\, (k-\widetilde{k})\mathrm{mod}K\right)
\end{aligned} \tag{A.3}$$

where $(a)$ follows from (24), $(b)$ relies on (25), and $(c)$ follows from (24) and (26). Similarly, for all $j, \widetilde{k} \in \{0, 1, \ldots, K-1\}$

$$
\begin{aligned}
\sum_{k'=0}^{K-1} \frac{1}{K} P(g_{\widetilde{k}}(j)|k')^{1-s} & \overset{(a)}{=} \sum_{k'=0}^{K-1} \frac{1}{K} P\big(j|\,(k'-\widetilde{k})\mathrm{mod}K\big)^{1-s} \\
& \overset{(b)}{=} \sum_{k'=0}^{K-1} \frac{1}{K} P(j|k')^{1-s}
\end{aligned}
\tag{A.4}
$$

where $(a)$ relies on (A.3) and $(b)$ holds since when $k'$ takes all the values in $\{0, 1, \ldots, K-1\}$, so does $(k'-\widetilde{k})\mathrm{mod}K$. Substituting (A.3) and (A.4) in (A.2) gives

$$
\begin{aligned}
& \sum_{j=0}^{J-1} \left\{ P(j|k)^{1-s} \left( \sum_{k'=0}^{K-1} \frac{1}{K} P(j|k')^{1-s} \right)^{\frac{s}{1-s}} \right\} \\
& = \sum_{\widetilde{k}=0}^{K-1} \sum_{j=0}^{J-1} \left\{ \frac{1}{K} P\big(j|(k-\widetilde{k})\mathrm{mod}K\big)^{1-s} \left( \sum_{k'=0}^{K-1} \frac{1}{K} P(j|k')^{1-s} \right)^{\frac{s}{1-s}} \right\} \\
& = \sum_{j=0}^{J-1} \left\{ \left( \sum_{\widetilde{k}=0}^{K-1} \frac{1}{K} P\big(j|(k-\widetilde{k})\mathrm{mod}K\big)^{1-s} \right) \left( \sum_{k'=0}^{K-1} \frac{1}{K} P(j|k')^{1-s} \right)^{\frac{s}{1-s}} \right\} \\
& \overset{(a)}{=} \sum_{j=0}^{J-1} \left\{ \left( \sum_{\widetilde{k}=0}^{K-1} \frac{1}{K} P(j|\widetilde{k})^{1-s} \right) \left( \sum_{k'=0}^{K-1} \frac{1}{K} P(j|k')^{1-s} \right)^{\frac{s}{1-s}} \right\} \\
& = \sum_{j=0}^{J-1} \left( \sum_{k'=0}^{K-1} \frac{1}{K} P(j|k')^{1-s} \right)^{\frac{1}{1-s}}
\end{aligned}
\tag{A.5}
$$

where equality $(a)$ holds since when the index $\widetilde{k}$ takes all the values in $\{0, 1, \ldots, K-1\}$, so does $(k-\widetilde{k})\mathrm{mod}K$. ∎

We now turn to explore how the symmetry of the channel and the input distribution $\mathbf{q}_s$ induce a symmetry on the probability tilting measure $f_s$.

    *b) On the symmetry of the tilting measure $f_s$ for strictly symmetric DMCs:*

    *Lemma A.2:* For all $s \in (0,1)$, $k \in \mathcal{K}$ and $j \in \mathcal{J}$, the tilting measure $f_s$ in (50) satisfies

$$
f_s(j) = f_s\big(g_k(j)\big)
\tag{A.6}
$$

    *Proof:* Examining the definition of $f_s$ in (50), it can be observed that it suffices to show that

$$
\alpha_{j,s} = \alpha_{g_k(j),s}, \quad \forall s \in (0,1), \ k \in \mathcal{K}, \ j \in \mathcal{J}
$$

where $\alpha_{j,s}$ is given in (49). This equality is proved in (A.4) referring to the uniform input distribution where $q_{k,s} = \frac{1}{K}$ for all $k \in \mathcal{K}$. ∎

Having established some symmetry properties of $\mathbf{q}_s$ and $f_s$, we are ready to prove equalities (51) – (53).

*c) On the independence of $\mu_k$ and its two derivatives from $k$:* As we have shown, the uniform distribution $\mathbf{q}_s$ satisfies (48) in equality for all inputs, so

$$
\begin{aligned}
\mu_k(s) &= \ln\left(\sum_j P(j|k)^{1-s} f_s(j)^s\right) \\
&\overset{(a)}{=} \ln\left(\sum_j P(j|k)^{1-s}\left(\alpha_{j,s}\right)^{\frac{s}{1-s}}\right) - s\ln\left(\sum_j \left(\alpha_{j,s}\right)^{\frac{1}{1-s}}\right) \\
&\overset{(b)}{=} (1-s)\ln\left(\sum_j \left(\alpha_{j,s}\right)^{\frac{1}{1-s}}\right) \\
&\overset{(c)}{=} (1-s)\ln\left(\sum_j \left[\sum_k q_{k,s} P(j|k)^{1-s}\right]^{\frac{1}{1-s}}\right)
\end{aligned}
\tag{A.7}
$$

where $(a)$ follows from the choice of $f_s$ in (49) and (50), $(b)$ follows from Lemma A.1 and (49), and $(c)$ follows from (49). Under the setting $s = \frac{\rho}{1+\rho}$, since the conditions on $\mathbf{q}_s$ in (48) are identical to the conditions on the input distribution $\mathbf{q} = \mathbf{q}_s$ which maximizes $E_0(\frac{s}{1-s}, \mathbf{q})$ as stated in [11, Theorem 4], then

$$
\begin{aligned}
\mu_k(s, f_s) &= (1-s)\ln\left(\sum_j \left[\sum_k q_{k,s} P(j|k)^{\frac{1}{1+\frac{s}{1-s}}}\right]^{1+\frac{s}{1-s}}\right) \\
&= -(1-s)\, E_0\left(\frac{s}{1-s}, \mathbf{q}_s\right) \\
&= -(1-s)\, E_0\left(\frac{s}{1-s}\right), \quad 0 < s < 1
\end{aligned}
\tag{A.8}
$$

where $E_0$ is given in (19). This proves (51).

We now turn to prove the independence of the first two derivatives of $\mu_k$ w.r.t $s$ from $k \in \mathcal{K}$.

*Remark A.1:* Note that the partial derivative of $\mu_k(s)$ w.r.t $s$ is performed while holding $f = f_s$ constant.

As is shown in [25],

$$
\begin{aligned}
\mu'(s) &= \mathbb{E}_{Q_s}\big(D(j)\big) \\
\mu''(s) &= \text{Var}_{Q_s}\big(D(j)\big)
\end{aligned}
$$

where

$$
D(j) \triangleq \ln\left(\frac{P_2(j)}{P_1(j)}\right), \quad Q_s(j) \triangleq \frac{P_1(j)^{1-s} P_2(j)^s}{\sum_{j'} P_1(j')^{1-s} P_2(j')^s}.
$$

For every $k \in \mathcal{K}$, $P_1$ and $P_2$ used in $\mu_k$ are defined to be $P(\cdot|k)$ and $f_s$, respectively. Hence, for all $k \in \mathcal{K}$

$$
\begin{aligned}
\mu_k'(s, f_s) &= \mathbb{E}_{Q_{k,s}}\big(D_{k,s}(j)\big) \\
\mu_k''(s, f_s) &= \text{Var}_{Q_{k,s}}\big(D_{k,s}(j)\big)
\end{aligned}
\tag{A.9}
$$

where

$$
D_{k,s}(j) \triangleq \ln\left(\frac{f_s(j)}{P(j|k)}\right)
\tag{A.10}
$$

$$
Q_{k,s}(j) \triangleq \frac{P(j|k)^{1-s} f_s(j)^s}{\sum_{j'=0}^{J-1} P(j'|k)^{1-s} f_s(j')^s}.
\tag{A.11}
$$

Applying (24) and Lemma A.2, we get that for all $k \in \mathcal{K}$

$$
\sum_{j'=0}^{J-1} P(j'|k)^{1-s} f_s(j')^s \overset{(a)}{=} \sum_{j'=0}^{J-1} P(g_k^{-1}(j')|0)^{1-s} f_s(g_k^{-1}(j'))^s
$$

$$
\overset{(b)}{=} \sum_{j'=0}^{J-1} P(j'|0)^{1-s} f_s(j')^s \tag{A.12}
$$

where $(a)$ follows from (24) and Lemma A.2 and $(b)$ follows since $g_k^{-1}$ is unitary. Substituting (A.12) in (A.11) gives

$$
\begin{aligned}
Q_{k,s}(j) &= \frac{P(j|k)^{1-s} f_s(j)^s}{\displaystyle\sum_{j'=0}^{J-1} P(j'|k)^{1-s} f_s(j')^s} \\
&\overset{(a)}{=} \frac{P(g_k^{-1}(j)|0)^{1-s} f_s(g_k^{-1}(j))^s}{\displaystyle\sum_{j'=0}^{J-1} P(j'|0)^{1-s} f_s(j')^s} \\
&\overset{(b)}{=} Q_{0,s}(g_k^{-1}(j)) \tag{A.13}
\end{aligned}
$$

where $(a)$ follows from (24), (26), (50), (A.6) and (A.12), and $(b)$ relies on the definition of $Q_{k,s}$ in (A.11). Similarly,

$$
\begin{aligned}
D_{k,s}(j) &= \ln\left(\frac{f_s(j)}{P(j|k)}\right) \\
&\overset{(a)}{=} \ln\left(\frac{f_s(g_k^{-1}(j))}{P(g_k^{-1}(j)|0)}\right) \\
&\overset{(b)}{=} D_{0,s}(g_k^{-1}(j)) \tag{A.14}
\end{aligned}
$$

where $(a)$ follows from (24), (26) and (A.6), and $(b)$ relies on the definition of $D_{k,s}$ in (A.10). Using (A.13) and (A.14), we finally get for all $k \in \mathcal{K}$

$$
\begin{aligned}
\mu_k'(s) &= \mathbb{E}_{Q_{k,s}}(D_{k,s}(j)) \\
&= \sum_{j=0}^{J-1} Q_{k,s}(j) D_{k,s}(j) \\
&= \sum_{j=0}^{J-1} Q_{0,s}(g_k^{-1}(j)) D_{0,s}(g_k^{-1}(j)) \\
&\overset{(a)}{=} \sum_{j=0}^{J-1} Q_{0,s}(j) D_{0,s}(j) \\
&= \mu_0'(s)
\end{aligned}
$$

and

$$
\begin{aligned}
\mu_k''(s) &= \operatorname{Var}_{Q_{k,s}}\big(D_{k,s}(j)\big) \\
&= \sum_{j=0}^{J-1} Q_{k,s}(j) D_{k,s}^2(j) - \mu_k'(s)^2 \\
&= \sum_{j=0}^{J-1} Q_{0,s}\big(g_k^{-1}(j)\big)\Big(D_{0,s}\big(g_k^{-1}(j)\big)\Big)^2 - \mu_0'(s)^2 \\
&\overset{(b)}{=} \sum_{j=0}^{J-1} Q_{0,s}(j)\big(D_{0,s}(j)\big)^2 - \mu_0'(s)^2 \\
&= \mu_0''(s)
\end{aligned}
$$

where $(a)$ and $(b)$ follow since $g_k^{-1}$ is unitary for all $k \in \mathcal{K}$. This completes the proof of Lemma 3.1.

*Remark A.2:* Equalities (51)–(53) hold for arbitrary symmetric memoryless channels. For a general output alphabet $\mathcal{J} \subseteq \mathbb{R}^d$, the proof of these properties follows the same lines as the proof here with the exception that the sums over $\mathcal{J}$ are replaced by integrals. As in Definition 3.1, if the projection of $\mathcal{J}$ over some of the $d$ dimensions is countable, the integration over these dimensions is turned into a sum.

## APPENDIX B: CALCULATION OF THE FUNCTION $\mu_0$ IN (47) FOR SOME SYMMETRIC CHANNELS

This appendix presents some technical calculations which yield the expressions for the function $\mu_0$ defined in (47) and its first two derivatives w.r.t. $s$ (while holding $f_s$ fixed in the calculation of the partial derivatives of $\mu$ w.r.t. $s$, as required in [25]). The examined cases are M-ary PSK modulated signals transmitted over an AWGN channel and binary block codes transmitted over the BEC. These expressions serve for the application of the VF bound in [35] and the ISP bound derived in Section III to block codes transmitted over these channels.

### A. The M-ary PSK modulated AWGN channel

For M-ary PSK modulated signals transmitted over the AWGN channel, the channel output is $\mathcal{J} = \mathbb{R}^2$. In the case of a continuous output alphabet, the sums in (A.7) are replaced by integrals, and the transition probabilities are replaced by transition probability density functions. Due to the symmetry of the channel, we get from Lemma A.1 that the distribution $\mathbf{q}_s$ which satisfies (48) is uniform. Hence, we get by substituting (80) into (A.7) that

$$
\mu_0(s) = (1-s)\ln\left(\iint_{\mathbb{R}^2} \frac{1}{2\pi\sigma^2} e^{-\frac{\|\mathbf{y}-\mathbf{x}_0\|^2}{2\sigma^2}} \left(\frac{1}{M}\sum_{k=0}^{M-1} e^{-\frac{(1-s)\,(\|\mathbf{y}-\mathbf{x}_k\|^2 - \|\mathbf{y}-\mathbf{x}_0\|^2)}{2\sigma^2}}\right)^{\frac{1}{1-s}} d\mathbf{y}\right).
$$

Since $\|x_k\|^2 = 1$ for all $k \in \{0, 1, \ldots, M-1\}$ we have

$$
\|\mathbf{y}-\mathbf{x}_k\|^2 - \|\mathbf{y}-\mathbf{x}_0\|^2 = -2\langle \mathbf{y}, \mathbf{x}_k - \mathbf{x}_0\rangle \tag{B.1}
$$

and so $\mu_0$ can be rewritten in the form $\mu_0(s) = (1-s)\ln\big(\theta(s)\big)$ where

$$
\theta(s) \triangleq \iint_{\mathbb{R}^2} \frac{1}{2\pi\sigma^2} e^{-\frac{\|\mathbf{y}-\mathbf{x}_0\|^2}{2\sigma^2}} \left(\frac{1}{M}\sum_{k=0}^{M-1} e^{\frac{(1-s)\,\langle \mathbf{y},\mathbf{x}_k-\mathbf{x}_0\rangle}{\sigma^2}}\right)^{\frac{1}{1-s}} d\mathbf{y}. \tag{B.2}
$$

We now turn to calculate the derivative of $\mu_0$ with respect to $s$ while holding $f = f_s$ constant. Substituting (80)

into the definition of $f_s$ in (50), we get that $f_s$ is given by

$$
\begin{aligned}
f_s(\mathbf{y}) &= \frac{\left(\sum_{k=0}^{M-1} \frac{1}{M}\left(\frac{1}{2\pi\sigma^2}\right)^{1-s} e^{-\frac{(1-s)\|\mathbf{y}-\mathbf{x}_k\|^2}{2\sigma^2}}\right)^{\frac{1}{1-s}}}{\iint_{\mathbb{R}^2}\left(\sum_{k=0}^{M-1} \frac{1}{M}\left(\frac{1}{2\pi\sigma^2}\right)^{1-s} e^{-\frac{(1-s)\|\mathbf{y}'-\mathbf{x}_k\|^2}{2\sigma^2}}\right)^{\frac{1}{1-s}} d\mathbf{y}'} \\
&= \frac{1}{2\pi\sigma^2}\frac{1}{\theta(s)}\cdot e^{-\frac{\|\mathbf{y}-\mathbf{x}_0\|^2}{2\sigma^2}}\left(\frac{1}{M}\sum_{k=0}^{M-1} e^{\frac{(1-s)\langle\mathbf{y},\mathbf{x}_k-\mathbf{x}_0\rangle}{\sigma^2}}\right)^{\frac{1}{1-s}}
\end{aligned} \tag{B.3}
$$

where the last equality follows from (B.1) and (B.2). The log-likelihood ratio $D_{0,s}$ in (A.10) is given by

$$
\begin{aligned}
D_{0,s}(\mathbf{y}) &\triangleq \ln\left(\frac{f_s(\mathbf{y})}{P(\mathbf{y}|0)}\right) \\
&= \frac{1}{1-s}\ln\left(\frac{1}{M}\sum_{k=0}^{M-1} e^{\frac{(1-s)\langle\mathbf{y},\mathbf{x}_k-\mathbf{x}_0\rangle}{\sigma^2}}\right) - \ln\left(\theta(s)\right)
\end{aligned} \tag{B.4}
$$

where the second equality follows from (80) and (B.3). The distribution $Q_{0,s}$ in (A.11) is given by

$$
\begin{aligned}
Q_{0,s}(\mathbf{y}) &\triangleq \frac{P(\mathbf{y}|0)^{1-s}f_s(\mathbf{y})^s}{\iint_{\mathbb{R}^2} P(\mathbf{y}'|0)^{1-s}f_s(\mathbf{y}')^s\, d\mathbf{y}'} \\[2mm]
&= \frac{\frac{1}{2\pi\sigma^2}e^{-\frac{\|\mathbf{y}-\mathbf{x}_0\|^2}{2\sigma^2}}\left(\frac{1}{M}\sum_{k=0}^{M-1} e^{\frac{(1-s)\langle\mathbf{y},\mathbf{x}_k-\mathbf{x}_0\rangle}{\sigma^2}}\right)^{\frac{s}{1-s}}}{\iint_{\mathbb{R}^2}\frac{1}{2\pi\sigma^2}e^{-\frac{\|\mathbf{y}'-\mathbf{x}_0\|^2}{2\sigma^2}}\left(\frac{1}{M}\sum_{k=0}^{M-1} e^{\frac{(1-s)\langle\mathbf{y}',\mathbf{x}_k-\mathbf{x}_0\rangle}{\sigma^2}}\right)^{\frac{s}{1-s}} d\mathbf{y}'} \\[2mm]
&\stackrel{(a)}{=} \frac{\frac{1}{2\pi\sigma^2}e^{-\frac{\|\mathbf{y}-\mathbf{x}_0\|^2}{2\sigma^2}}\left(\frac{1}{M}\sum_{k=0}^{M-1} e^{\frac{(1-s)\langle\mathbf{y},\mathbf{x}_k-\mathbf{x}_0\rangle}{\sigma^2}}\right)^{\frac{s}{1-s}}}{\iint_{\mathbb{R}^2}\left(\frac{1}{2\pi\sigma^2}e^{-\frac{\|\mathbf{y}'-\mathbf{x}_0\|^2}{2\sigma^2}}\right)^{1-s}\left(\frac{1}{M}\sum_{k=0}^{M-1}\left(\frac{1}{2\pi\sigma^2}e^{-\frac{\|\mathbf{y}'-\mathbf{x}_k\|^2}{2\sigma^2}}\right)^{1-s}\right)^{\frac{s}{1-s}} d\mathbf{y}'} \\[2mm]
&\stackrel{(b)}{=} \frac{\frac{1}{2\pi\sigma^2}e^{-\frac{\|\mathbf{y}-\mathbf{x}_0\|^2}{2\sigma^2}}\left(\frac{1}{M}\sum_{k=0}^{M-1} e^{\frac{(1-s)\langle\mathbf{y},\mathbf{x}_k-\mathbf{x}_0\rangle}{\sigma^2}}\right)^{\frac{s}{1-s}}}{\iint_{\mathbb{R}^2}\left(\frac{1}{M}\sum_{k=0}^{M-1}\left(\frac{1}{2\pi\sigma^2}e^{-\frac{\|\mathbf{y}'-\mathbf{x}_k\|^2}{2\sigma^2}}\right)^{1-s}\right)^{\frac{1}{1-s}} d\mathbf{y}'} \\[2mm]
&\stackrel{(c)}{=} \frac{\frac{1}{2\pi\sigma^2}e^{-\frac{\|\mathbf{y}-\mathbf{x}_0\|^2}{2\sigma^2}}\left(\frac{1}{M}\sum_{k=0}^{M-1} e^{\frac{(1-s)\langle\mathbf{y},\mathbf{x}_k-\mathbf{x}_0\rangle}{\sigma^2}}\right)^{\frac{s}{1-s}}}{\iint_{\mathbb{R}^2}\frac{1}{2\pi\sigma^2}e^{-\frac{\|\mathbf{y}'-\mathbf{x}_0\|^2}{2\sigma^2}}\left(\frac{1}{M}\sum_{k=0}^{M-1} e^{\frac{(1-s)\langle\mathbf{y}',\mathbf{x}_k-\mathbf{x}_0\rangle}{\sigma^2}}\right)^{\frac{1}{1-s}} d\mathbf{y}'} \\[2mm]
&\stackrel{(d)}{=} \frac{1}{2\pi\sigma^2\,\theta(s)}\, e^{-\frac{\|\mathbf{y}-\mathbf{x}_0\|^2}{2\sigma^2}}\left(\frac{1}{M}\sum_{k=0}^{M-1} e^{\frac{(1-s)\langle\mathbf{y},\mathbf{x}_k-\mathbf{x}_0\rangle}{\sigma^2}}\right)^{\frac{s}{1-s}}
\end{aligned} \tag{B.5}
$$

where $(a)$ and $(c)$ rely on (B.1), $(b)$ follows from Lemma 2.1 in the proof for symmetric channels, and $(d)$ relies on the definition of $\theta$ in (B.2). Substituting (B.4) and (B.5) in (A.9) we get

$$
\begin{aligned}
\mu_0'(s) &= \mathbb{E}_{Q_{0,s}}\left(D_{0,s}\right) \\
&= \frac{1}{(1-s)\theta(s)} \iint_{\mathbb{R}^2} \frac{1}{2\pi\sigma^2} e^{-\frac{\|\mathbf{y}-\mathbf{x}_0\|^2}{2\sigma^2}} \left(\frac{1}{M}\sum_{k=0}^{M-1} e^{\frac{(1-s)\,\langle\mathbf{y},\mathbf{x}_k-\mathbf{x}_0\rangle}{\sigma^2}}\right)^{\frac{s}{1-s}} \\
&\qquad\qquad\qquad \cdot \ln\left(\frac{1}{M}\sum_{k=0}^{M-1} e^{\frac{(1-s)\,\langle\mathbf{y},\mathbf{x}_k-\mathbf{x}_0\rangle}{\sigma^2}}\right) d\mathbf{y} - \ln\left(\theta(s)\right)
\end{aligned}
\tag{B.6}
$$

and

$$
\begin{aligned}
\mu_0''(s) &= \mathbb{E}_{Q_{0,s}}\left(D_{0,s}^2(\mathbf{y})\right) - \mu_0'(s)^2 \\
&= \frac{1}{\theta(s)} \iint_{\mathbb{R}^2} \frac{1}{2\pi\sigma^2} e^{-\frac{\|\mathbf{y}-\mathbf{x}_0\|^2}{2\sigma^2}} \left(\frac{1}{M}\sum_{k=0}^{M-1} e^{\frac{(1-s)\,\langle\mathbf{y},\mathbf{x}_k-\mathbf{x}_0\rangle}{\sigma^2}}\right)^{\frac{s}{1-s}} \\
&\qquad \cdot \left(\frac{1}{1-s}\ln\left(\frac{1}{M}\sum_{k=0}^{M-1} e^{\frac{(1-s)\,\langle\mathbf{y},\mathbf{x}_k-\mathbf{x}_0\rangle}{\sigma^2}}\right) - \ln\left(\theta(s)\right)\right)^2 d\mathbf{y} - \mu_0'(s)^2.
\end{aligned}
\tag{B.7}
$$

## B. The Binary Erasure Channel

Let us denote the output of the channel when an erasure has occurred by $\mathcal{E}$, and let $p$ designate the erasure probability of the channel. Since the BEC is symmetric, the input distribution $\mathbf{q}_s$ which satisfies (48) is uniform (see Lemma A.1), and we get from (A.7)

$$
\begin{aligned}
\mu_0(s, f_s) &= (1-s)\ln\left(\frac{2(1-p)}{2^{\frac{1}{1-s}}} + p\right) \\
&= (1-s)\ln\left(2(1-p) + 2^{\frac{1}{1-s}}p\right) - \ln 2.
\end{aligned}
\tag{B.8}
$$

We now turn to calculate $f_s$ for the BEC; substituting the transition probabilities into (50) gives

$$
\begin{aligned}
f_s(0) = f_s(1) &= \frac{\left(\frac{1}{2}(1-p)^{1-s}\right)^{\frac{1}{1-s}}}{2\left(\frac{1}{2}(1-p)^{1-s}\right)^{\frac{1}{1-s}} + (p^{1-s})^{\frac{1}{1-s}}} \\
&= \frac{2^{-\frac{1}{1-s}}(1-p)}{2^{1-\frac{1}{1-s}}(1-p) + p} \\
&= \frac{1-p}{2(1-p) + 2^{\frac{1}{1-s}}p}
\end{aligned}
\tag{B.9}
$$

and

$$
\begin{aligned}
f_s(\mathcal{E}) &= \frac{\left(p^{1-s}\right)^{\frac{1}{1-s}}}{2\left(\frac{1}{2}(1-p)^{1-s}\right)^{\frac{1}{1-s}} + (p^{1-s})^{\frac{1}{1-s}}} \\
&= \frac{p}{2^{1-\frac{1}{1-s}}(1-p) + p} \\
&= \frac{2^{\frac{1}{1-s}}p}{2(1-p) + 2^{\frac{1}{1-s}}p}.
\end{aligned}
\tag{B.10}
$$

Substituting (B.9) and (B.10) into the definition of the distribution $Q_{0,s}$ in (A.11) gives

$$
\begin{aligned}
Q_{0,s}(0) &= \frac{P(0|0)^{1-s} f_s(0)^s}{P(0|0)^{1-s} f_s(0)^s + P(1|0)^{1-s} f_s(1)^s + P(\mathcal{E}|0)^{1-s} f_s(\mathcal{E})^s} \\
&= \frac{1-p}{1-p+2^{\frac{s}{1-s}}p} \\
Q_{0,s}(1) &= \frac{P(1|0)^{1-s} f_s(0)^s}{P(0|0)^{1-s} f_s(0)^s + P(1|0)^{1-s} f_s(1)^s + P(\mathcal{E}|0)^{1-s} f_s(\mathcal{E})^s} \\
&= 0 \\
Q_{0,s}(\mathcal{E}) &= \frac{P(\mathcal{E}|0)^{1-s} f_s(\mathcal{E})^s}{P(0|0)^{1-s} f_s(0)^s + P(1|0)^{1-s} f_s(1)^s + P(\mathcal{E}|0)^{1-s} f_s(\mathcal{E})^s} \\
&= \frac{2^{\frac{s}{1-s}}p}{1-p+2^{\frac{s}{1-s}}p}
\end{aligned}
\tag{B.11}
$$

and the LLR in (A.10) is given by

$$
\begin{aligned}
D_{0,s}(0) &= \ln\left(\frac{1}{2(1-p)+2^{\frac{1}{1-s}}p}\right) \\
D_{0,s}(\mathcal{E}) &= \ln\left(\frac{2^{\frac{1}{1-s}}}{2(1-p)+2^{\frac{1}{1-s}}p}\right).
\end{aligned}
\tag{B.12}
$$

Applying (B.11) and (B.12) we get from (A.11)

$$
\begin{aligned}
\mu_0'(s,f_s) &= \mathbb{E}_{Q_{0,s}}(D_{0,s}) \\
&= \frac{1-p}{1-p+2^{\frac{s}{1-s}}p}\ln\left(\frac{1}{2(1-p)+2^{\frac{1}{1-s}}p}\right) \\
&\quad + \frac{2^{\frac{s}{1-s}}p}{1-p+2^{\frac{s}{1-s}}p}\ln\left(\frac{2^{\frac{1}{1-s}}}{2(1-p)+2^{\frac{1}{1-s}}p}\right) \\
&= \ln\left(\frac{1}{1-p+2^{\frac{s}{1-s}}p}\right) + \frac{2^{\frac{s}{1-s}}p}{1-p+2^{\frac{s}{1-s}}p}\frac{\ln 2}{1-s}
\end{aligned}
\tag{B.13}
$$

and

$$
\begin{aligned}
\mu_0''(s,f_s) &= \mathbb{E}_{Q_{0,s}}(D_{0,s}^2) - \mu_0'(s,f_s)^2 \\
&= \frac{1-p}{1-p+2^{\frac{s}{1-s}}p}\ln^2\left(\frac{1}{2(1-p)+2^{\frac{1}{1-s}}p}\right) \\
&\quad + \frac{2^{\frac{s}{1-s}}p}{1-p+2^{\frac{s}{1-s}}p}\ln^2\left(\frac{2^{\frac{1}{1-s}}}{2(1-p)+2^{\frac{1}{1-s}}p}\right) - \mu_0'(s,f_s)^2 \\
&= \ln^2\left(\frac{1}{1-p+2^{\frac{s}{1-s}}p}\right) + \frac{2^{\frac{1}{1-s}}p}{1-p+2^{\frac{s}{1-s}}p}\frac{\ln 2}{1-s}\ln\left(\frac{1}{1-p+2^{\frac{s}{1-s}}p}\right) \\
&\quad + \frac{2^{\frac{s}{1-s}}p}{1-p+2^{\frac{s}{1-s}}p}\left(\frac{\ln 2}{1-s}\right)^2 - \mu_0'(s,f_s)^2 \\
&= \frac{2^{\frac{s}{1-s}}p(1-p)}{\left(1-p+2^{\frac{s}{1-s}}p\right)^2}\left(\frac{\ln 2}{1-s}\right)^2
\end{aligned}
\tag{B.14}
$$

## Appendix C: Proof of Proposition 4.2

From the definition of $f_N$ in (67), it follows that

$$
\begin{aligned}
f_N(x) &= \frac{1}{2^{\frac{N-1}{2}}\Gamma(\frac{N+1}{2})} \int_0^\infty z^{N-1} \exp(-\frac{z^2}{2} + zx)\, dz \\
&= \frac{e^{\frac{x^2}{2}}}{2^{\frac{N-1}{2}}\Gamma(\frac{N+1}{2})} \int_0^\infty z^{N-1} \exp\left(-\frac{(z-x)^2}{2}\right)\, dz \\
&= \frac{e^{\frac{x^2}{2}}}{2^{\frac{N-1}{2}}\Gamma(\frac{N+1}{2})} \int_{-x}^\infty (u+x)^{N-1} \exp\left(-\frac{u^2}{2}\right)\, du \ .
\end{aligned}
$$

From the binomial formula, we get

$$
f_N(x) = \frac{e^{\frac{x^2}{2}}}{2^{\frac{N-1}{2}}\Gamma(\frac{N+1}{2})} \sum_{j=0}^{N-1} \left[ \binom{N-1}{j} x^{N-1-j} \int_{-x}^\infty u^j \exp\left(-\frac{u^2}{2}\right)\, du \right] \ . \tag{C.1}
$$

We now examine the integrals in the RHS of (C.1). For odd values of $j$, we get

$$
\begin{aligned}
\int_{-x}^\infty u^j \exp\left(-\frac{u^2}{2}\right)\, du &= \int_{-x}^x u^j \exp\left(-\frac{u^2}{2}\right)\, du + \int_x^\infty u^j \exp\left(-\frac{u^2}{2}\right)\, du \\
&= \int_x^\infty u^j \exp\left(-\frac{u^2}{2}\right)\, du \\
&= \int_0^\infty u^j \exp\left(-\frac{u^2}{2}\right)\, du - \int_0^x u^j \exp\left(-\frac{u^2}{2}\right)\, du \tag{C.2}
\end{aligned}
$$

where the second equality follows since the integrand is an odd function and the domain of first integral is symmetric around zero. For even values of $j$, we get

$$
\begin{aligned}
\int_{-x}^\infty u^j \exp\left(-\frac{u^2}{2}\right)\, du &= \int_0^\infty u^j \exp\left(-\frac{u^2}{2}\right)\, du + \int_{-x}^0 u^j \exp\left(-\frac{u^2}{2}\right)\, du \\
&= \int_0^\infty u^j \exp\left(-\frac{u^2}{2}\right)\, du + \int_0^x u^j \exp\left(-\frac{u^2}{2}\right)\, du \tag{C.3}
\end{aligned}
$$

where the second equality holds since the integrand is an even function. Combining (C.2) and (C.3) gives that for $j \in \{0, 1, \ldots, N-1\}$

$$
\begin{aligned}
\int_{-x}^\infty u^j \exp\left(-\frac{u^2}{2}\right)\, du &= \int_0^\infty u^j \exp\left(-\frac{u^2}{2}\right)\, du + (-1)^j \int_0^x u^j \exp\left(-\frac{u^2}{2}\right)\, du \\
&\stackrel{(a)}{=} \int_0^\infty (2t)^{\frac{j-1}{2}} e^{-t}\, dt + (-1)^j \int_0^{\frac{x^2}{2}} (2t)^{\frac{j-1}{2}} e^{-t}\, dt \\
&= 2^{\frac{j-1}{2}} \int_0^\infty t^{\frac{j-1}{2}} e^{-t}\, dt \left[ 1 + (-1)^j \frac{\displaystyle\int_0^{\frac{x^2}{2}} t^{\frac{j-1}{2}} e^{-t}\, dt}{\displaystyle\int_0^\infty t^{\frac{j-1}{2}} e^{-t}\, dt} \right] \\
&= 2^{\frac{j-1}{2}} \Gamma\left(\frac{j+1}{2}\right) \left[ 1 + (-1)^j\, \tilde{\gamma}\left(\frac{x^2}{2}, \frac{j+1}{2}\right) \right]
\end{aligned}
$$

where $(a)$ follows by substituting $t \triangleq \frac{u^2}{2}$ and the functions $\Gamma$ and $\tilde{\gamma}$ are defined in (75) and (76), respectively. Substituting the last equality in (C.1) and also noting that

$$
\binom{N-1}{j} = \frac{\Gamma(N)}{\Gamma(N-j)\,\Gamma(j+1)} \ , \quad N \in \mathbb{N},\ j \in \{0, 1, \ldots, N-1\}
$$

we get

$$
\begin{aligned}
f_N(x) &= \frac{e^{\frac{x^2}{2}}}{2^{\frac{N-1}{2}}\Gamma(\frac{N+1}{2})} \sum_{j=0}^{N-1} \left\{ \frac{\Gamma(N)}{\Gamma(N-j)\,\Gamma(j+1)} \; x^{N-1-j} \; 2^{\frac{j-1}{2}} \right. \\
&\qquad\qquad\qquad\qquad \left. \cdot\, \Gamma\left(\frac{j+1}{2}\right) \left[ 1+(-1)^j \, \tilde{\gamma}\left(\frac{x^2}{2},\frac{j+1}{2}\right) \right] \right\} \\
&= \sum_{j=0}^{N-1} \left\{ \frac{e^{\frac{x^2}{2}}}{\Gamma(N-j)} \; \frac{\Gamma(N)}{\Gamma\left(\frac{N+1}{2}\right)} \; \frac{\Gamma\left(\frac{j+1}{2}\right)}{\Gamma(j+1)} \; \frac{x^{N-1-j}}{2^{\frac{N-j}{2}}} \left[ 1+(-1)^j \, \tilde{\gamma}\left(\frac{x^2}{2},\frac{j+1}{2}\right) \right] \right\} \\
&\overset{(a)}{=} \sum_{j=0}^{N-1} \left\{ \frac{e^{\frac{x^2}{2}}}{\Gamma(N-j)} \; \frac{2^{N-1}\,\Gamma\left(\frac{N}{2}\right)}{\sqrt{\pi}} \; \frac{2^{-j}\,\sqrt{\pi}}{\Gamma\left(\frac{j}{2}+1\right)} \; \frac{x^{N-1-j}}{2^{\frac{N-j}{2}}} \left[ 1+(-1)^j \, \tilde{\gamma}\left(\frac{x^2}{2},\frac{j+1}{2}\right) \right] \right\} \\
&\overset{(b)}{=} \sum_{j=0}^{N-1} \exp\big( d(N,j,x) \big)
\end{aligned}
$$

where $(a)$ follows from the equality

$$
\Gamma(2u) = \frac{2^{2u-1}}{\sqrt{\pi}} \, \Gamma(u) \, \Gamma\left( u+\frac{1}{2}\right), \quad u \neq 0, -\frac{1}{2}, -1, -\frac{3}{2}, \dots
$$

and $(b)$ follows from the definition of $d(N,j,x)$ in (74).

## REFERENCES

[1] B. Ammar, Y. Kou, J. Xu and S. Lin, "Construction of low-density parity-check codes based on balanced incomplete block designs," *IEEE Trans. on Information Theory*, vol. 50, no. 6, pp. 1257-1268, June 2004.

[2] E. R. Berlekamp, "The performance of block codes," Notices of the AMS, pp. 17–22, January 2002. [Online]. Available: `http://www.ams.org/notices/200201/fea-berlekamp.pdf`.

[3] D. J. Costello and G. D. Forney, "Channel coding: The road to channel capacity," submitted to the *Proceedings of the IEEE*, November 2006. [Online]. Available: `http://www.arxiv.org/abs/cs.IT/0611112`.

[4] J. Cuevas, P. Adde and S. Kerouedan, "Turbo decoding of product codes for Gigabit per second applications and beyond," *European Transactions on Telecommunications*, vol. 17, no. 1, pp. 45–55, Jan.–Feb. 2006.

[5] C. Di, D. Proietti, I. E. Telatar and R. Urbanke, "Finite-length analysis of low-density parity-check codes," *IEEE Trans. on Information Theory*, vol. 48, no. 6, pp. 1570–1579, June 2002.

[6] D. Divsalar and S. Doliner, "Concatenation of Hamming codes and accumulator codes with high-order modulations for high-speed decoding," Jet Propulsion Laboratory (JPL), IPN Progress Report 42-156, February 15, 2004.
[Online]. Available: `http://tmo.jpl.nasa.gov/progress_report/42-156/156G.pdf`.

[7] D. Divsalar and C. Jones, "Protograph LDPC codes with node degrees at least 3," *Proceedings of the 2006 IEEE Global Communications Conference (GlobeCom)*, San Francisco, CA, USA, 27 November–1 December 2006.

[8] D. Divsalar, C. Jones, S. Doliner and J. Thorpe, "Protograph based LDPC codes with minimum distance linearly growing with block size" *Proceedings of the 2005 IEEE Global Communications Conference (GlobeCom)*, vol. 3, pp. 1152–1156, St. Louis, MO, USA, 28 November–2 December 2005.

[9] S. Doliner, D. Divsalar and F. Pollara, "Code performance as a function of block size," Jet Propulsion Laboratory (JPL), TMO Progress Report 42-133, May 15, 1998.
[Online]. Available: `http://tmo.jpl.nasa.gov/tmo/progress_report/42-133/133K.pdf`.

[10] P. Elias, "List decoding for noisy channels," MIT Res. Lab. Electron., Cambridge, MA, USA, September 1957.

[11] R. G. Gallager, "A simple derivation of the coding theorem and some applications," *IEEE Trans. on Information Theory*, vol. 11, pp. 3–18, January 1965.

[12] R. G. Gallager, *Information Theory and Reliable Communications*, John Wiley, 1968.

[13] H. Herzberg and G. Poltyrev, "The error probability of M-ary PSK block coded modulation schemes," *IEEE Trans. on Communications*, vol. 44, no. 4, pp. 427–433, April 1996.

[14] D. E. Lazic, Th. Beth and M. Calic, "How close are turbo codes to optimal codes ?," *Proceedings of the International Symposium on Turbo Codes and Related Topics*, pp. 192–195, Brest, France, 3–5 September 1997.

[15] D. E. Lazic, Th. Beth and S. Egner, "Constrained capacity of the AWGN channel," *IEEE 1998 International Symposium on Information Theory (ISIT 1998)*, p. 237, Cambridge, MA, USA, 16–21 August, 1998.

[16] M. G. Luby, M. Mitzenmacher, M. A. Shokrollahi and D. A. Spielman, "Efficient erasure correcting codes," *IEEE Trans. on Information Theory*, vol. 47, no. 2, pp. 569–584, February 2001.

[17] S. J. Macmullan and O. M. Collins, "A comparison of known codes, random codes and the best codes," *IEEE Trans. on Information Theory*, vol. 44, no. 7, pp. 3009–3022, November 1998.

[18] H. D. Pfister, I. Sason, and R. Urbanke, "Capacity-achieving ensembles for the binary erasure channel with bounded complexity," *IEEE Trans. on Information Theory*, vol. 51, no. 7, pp. 2352–2379, July 2005.

[19] H. Pfister and I. Sason, "Accumulate-repeat-accumulate codes: Capacity-achieving ensembles of systematic codes for the erasure channel with bounded complexity," to appear in the *IEEE Trans. on Information Theory*, vol. 53, no. 6, June 2007.

[20] G. Poltyrev, "Bounds on the decoding error probability of binary linear codes via their spectra," *IEEE Trans. on Information Theory*, vol. 40, no. 4, pp. 1284–1292, July 1994.

[21] T. Richardson and R. Urbanke, *Modern Coding Theory*, to be published in 2007, Cambridge Press.
[Online]. Available: `http://lthcwww.epfl.ch/mct/index.php`.

[22] I. Sason and R. Urbanke, "Parity-check density versus performance of binary linear block codes over memoryless symmetric channels," *IEEE Trans. on Information Theory*, vol. 49, no. 7, pp. 1611–1635, July 2003.

[23] I. Sason and S. Shamai, *Performance Analysis of Linear Codes under Maximum-Likelihood Decoding: A Tutorial*," *Foundations and Trends in Communications and Information Theory*, vol. 3, no. 1-2, pp. 1–222, NOW Publishers, Delft, the Netherlands, July 2006.
[Online]. Available: `http://www.ee.technion.ac.il/people/sason/monograph_postprint.pdf`.

[24] C. E. Shannon, "Probability of error for optimal codes in a Gaussian channel," *Bell System Technical Journal*, vol. 38, pp. 611–656, May 1959.

[25] C. Shannon, R. Gallager and E. Berlekamp, "Lower bounds to error probability for decoding on discrete memoryless channels," *Information and Control*, vol. 10, Part 1: pp. 65–103, and Part 2: pp. 522–552, February/May 1967.

[26] *Claude Elwood Shannon - Collected Papers*, edited by N. J. A. Sloane and A. D. Wyner, IEEE Press, 1993.

[27] A. Shokrollahi, "New sequences of time erasure codes approaching channel capacity," in *Proceedings of the 13th International Symposium on Applied Algebra, Algebraic Algorithms and Error-Correcting Codes*, Lectures Notes in Computer Science 1719, Springer Verlag, pp. 65–76, 1999.

[28] Y. Tai, L. Lan, L. Zeng, S. Lin and K. Abdel-Ghaffar, "Algebraic construction of quasi-cyclic LDPC codes for the AWGN and erasure channels," *IEEE Trans. on Communications*, vol. 54, no. 10, pp. 1756–1765, October 2006.

[29] O. Y. Takeshita, O. M. Collins, P. C. Massey and D. J. Costello, "On the frame-error rate of concatenated turbo codes," *IEEE Trans. on Communications*, vol. 49, no. 4, pp. 602–608, April 2001.

[30] H. Tang, J. Xu, S. Lin and K. Abdel-Ghaffar, "Codes on finite geometries," *IEEE Trans. on Information Theory*, vol. 51, no. 2, pp. 572–596, February 2005.

[31] M. Twitto, I. Sason and S. Shamai, "Tightened upper bounds on the ML decoding error probability of binary linear block codes," to appear in the *IEEE Trans. on Information Theory*, vol. 53, no. 4, April 2007.

[32] M. Twitto and I. Sason, "On the error exponents of some improved tangetial-sphere bound," *IEEE Trans. on Information Theory*, vol. 53, no. 3, pp. 1196–1210, March 2007.

[33] R. Urbanke, *Error floor calculator for the binary erasure channel*.
[Online]. Available: `http://lthcwww.epfl.ch/research/efc/`.

[34] A. Valembois and M. Fossorier, "Box and match techniques applied to soft-decision decoding," *IEEE Trans. on Information Theory*, vol. 50, no. 5, pp. 796–810, May 2004.

[35] A. Valembois and M. Fossorier, "Sphere-packing bounds revisited for moderate block length," *IEEE Trans. on Information Theory*, vol. 50, no. 12, pp. 2998–3014, Decemeber 2004.

[36] C. C. Wang, S. R. Kulkarni and H. V. Poor, "Finite-dimensional bounds on $\mathbb{Z}_m$ and binary LDPC codes with belief-propagation decoders," *IEEE Trans. on Information Theory*, vol. 53, no. 1, pp. 56–81, January 2007.

[37] L. Wei, "Near-optimum serial concatenation of single-parity codes with convolutional codes," *IEE Proceedings on Communications*, vol. 152, no. 4, pp. 397–403, August 2005.

[38] J. M. Wozencraft, "List decoding," MIT Research Lab. Electron., Cambridge, MA, USA, January 1958.